\documentclass[aps,prf,preprint,superscriptaddress]{revtex4-1}

\usepackage{amsmath}
\usepackage{graphicx}
\usepackage{bm}
\usepackage{xcolor}
\usepackage{algorithmic}
\usepackage{subfig}
\usepackage{verbatim}
\usepackage{wasysym}%
\usepackage[colorinlistoftodos]{todonotes}
\usepackage[colorlinks=true, allcolors=blue]{hyperref}
\usepackage[section]{placeins}
\usepackage[margin=2.5cm]{geometry}
\usepackage{tikz}
\usepackage{xcolor}
\usepackage{amssymb}
\newsavebox{\measurebox}
\usepackage{adjustbox}
\usepackage{floatrow}
\usepackage{subfig}
\usepackage{blindtext}
\usepackage{multirow}
\usepackage{subcaption}

\begin{document}
\title{Emergent coordination and propulsion of a model spherical ciliate}
\author{Hang Su}
\email{hang.su16@imperial.ac.uk}
\affiliation{%
Department of Mathematics, Imperial College London, SW7 2AZ, UK
}
\author{Timothy A Westwood}
\email{tim.westwood@fluidgravity.co.uk}
\affiliation{%
Department of Mathematics, Imperial College London, SW7 2AZ, UK
}
\affiliation{Fluid Gravity Engineering Ltd., Emsworth, Hampshire, PO10 7DX, UK}
\author{Eric E Keaveny}
\email{ekeaveny@imperial.ac.uk}
\affiliation{%
Department of Mathematics, Imperial College London, SW7 2AZ, UK
}

\begin{abstract}
A longstanding challenge in biofluid dynamics research is a mechanistic understanding of the coordinated movement of motile cilia and its resulting ability to facilitate fluid transport.  In this study, we develop numerical techniques to simultaneously compute the emergent coordination of and propulsion by filamentous model cilia covering the surface of a sphere.  To accomplish this, we develop what we refer to as the filament oscillator model, in which each cilium has two dynamic degrees of freedom: a phase variable that maps to a specific shape in a prescribed sequence, and an angle that describes the overall orientation of the sequence.  By varying a parameter related to cilium stiffness, we show that there is bistability between symplectic-like and diaplectic metachronal waves, provided that the stiffness is sufficiently low.  Above the critical stiffness, only diaplectic waves emerge.  Further, we analyse the propulsive capabilities and flow fields of the two emergent states, showing that diaplectic waves provide more efficient propulsion due to their shorter wavelengths.  In addition, we examine how introducing beat-plane tilt leads to ciliate rotation while maintaining nearly identical emergent states and comparable swimming speeds. 

\end{abstract}

\maketitle

\section{Introduction}
Motile cilia are slender, flexible organelles used by cells across eukaryotic life to move and manipulate the fluids that surround them \cite{brennen1977,gibbons1981,gilpin2020,Wan2024}. While many microscopic organisms, such as algae and protists \cite{goldstein2015,wan2024II} and larvae of marine invertebrates \cite{poon2025,Poon2023}, rely on cilia for propulsion in aquatic environments, the cilia in our own bodies play crucial roles in facilitating fluid transport in vital organs such as the brain \cite{faubel2016} and lungs \cite{sleigh1988}. Since the earliest observations of microscopic life \cite{van1800select}, ciliary motion and, in particular, their often stunning collective dynamics have long captivated scientists.  Their motion and resulting fluid flows provided inspiration for Taylor \cite{taylor1951} and Lighthill \cite{lighthill1952,Blake1971} in the 1950s to establish the swimming sheet and the squirmer models, respectively, that are now cornerstones of the field referred to as biofluid dynamics \cite{lighthill1975,childress1981,lauga2020}. 

Applying these models to cilia-driven propulsion, cilia motion is captured through an effective surface velocity, which, when coupled to force- and torque-free conditions, produces a net translation.  The spherical squirmer model \cite{lighthill1952,Blake1971,pedley2016II} in particular has been used widely to study the motion of ciliated microorganisms and since its inception, it has been extended to non-axisymmetric surface velocities \cite{pak2014} as well as non-spherical bodies \cite{keller1977}.  It has been used to study the interactions between swimming ciliates \cite{ishikawa2006,delmotte2015}, provide a framework with which to explore optimal propulsion and nutrient uptake \cite{michelin2010,michelin2011,michelin2013}, and to examine motility at finite Reynolds numbers \cite{wang2012,chisholm2016}, and in non-Newtonian media, such as viscoelastic fluids \cite{zhu2012,elfring2014,decorato2017}.  The squirmer model has been particularly useful in conjunction with experiments, assessing the swimming speed and rotation, nutrient uptake, and phototaxis of the spherical algal colony, \textit{Volvox} \cite{Drescher2009,goldstein2015,ishikawa2024}.  Additionally, the squirmer model has inspired the use of surface velocities to model other cilia-driven flows, including those in the lung, allowing for natural variations in cilia distribution and orientation to be explored \cite{ramirez2020}.  In all of its variants, the squirmer model, and surface-flow models more generally, require the fluid velocity at the effective surface to be prescribed.  As a result, they do not describe how the surface flow arises, and extracting the surface velocity directly from ciliary motion has proven challenging \cite{pedley2016squirmers}.  Furthermore, recent simulations of ciliates \cite{Omori2020,Ito2019} have shown that while the squirmer model accurately characterises swimming speeds, it does not accurately capture the overall viscous dissipation, which is instead dominated by cilia motion.  

Just as the squirmer model has formed the basis for understanding the fluid flows generated by cilia, the minimal rotor model \cite{golestanian2011hydrodynamic,lenz2006collective,niedermayer2008synchronization,brumley2012hydrodynamic,uchida2012hydrodynamic, hamilton2021changes,lagomarsino2003metachronal,wollin2011metachronal,kanale2022} has instead facilitated our understanding of cilia coupling through hydrodynamics.  In its most basic incarnation, the rotor model treats each cilium as a spherical particle moving along a prescribed path in the vicinity of a no-slip surface.  The flows generated by their motion provide the interactions between the rotors, coupling motion along their respective paths.   A desirable feature of this model is that there is a single degree of freedom, a dynamic phase variable, associated with each cilium, allowing for ease of computation, both numerically and analytically, as well as a clear connection with coupled oscillator dynamics.  Additionally, other features such as variable forcing \cite{meng2021conditions}, deformability of rotor path \cite{brumley2012hydrodynamic,Brumley2015}, and shape of the underlying surface \cite{nasouri2016,mannan2020} can be readily considered in the model.  The rotor model has been shown to exhibit a variety of collective dynamics, including metachronal waves \cite{niedermayer2008synchronization,meng2021conditions,Brumley2015,kanale2022} and hence allows for an exploration of the connection between the microscale parameters governing the rotor dynamics, such as path shape and orientation, and the resulting collective behaviour.  While the rotor model has provided a pathway to understanding collective cilia dynamics, a spherical particle does not accurately characterise the drag on a shape-changing, filamentous cilium, and further, the flow field in the vicinity of the cilium is not accurately captured by that generated by a single point force.  As a result, cilia-driven flow fields, a quantity readily captured by the squirmer model, will not necessarily be accurately described by the rotor model, and thus a precise connection between measured cilium beats, emergent states and overall function will require a more detailed modelling approach.   There has been concerted effort to overcome this limitation using filament-based models of cilia \cite{elgeti_emergence_2013,chakrabarti2021multiscale,guirao2007spontaneous,gueron_cilia_1997,han2018spontaneous} to study emergent coordination, direct simulations at large-scale remain an ongoing challenge due both the fluid-structure interactions for deforming cilia and the lack of an accepted model for internal dynein forcing, though there has been recent progress in this area \cite{oriola2017nonlinear,chakrabarti2019spontaneous}.

In this paper, we close the gap between these different modelling paradigms and perform simulations of a model spherical ciliate whose motion is determined by the emergent coordination of filamentous cilia distributed over its surface, thus providing a direct link between emergent collective cilia dynamics and their hydrodynamic function.  To accomplish this, we build on the framework of the Lagrangian mechanics of active systems \cite{Solovev2021,solovev2022} to formulate what we refer to as the filament oscillator model.  In contrast to our previous work \cite{westwood2021coordinated} where cilia are treated as follower-force driven elastic filaments that experience beam-like elastic forces \cite{schoeller2021methods}, in the filament oscillator model, the cilia retain their filament-like shape, but cilia motion is determined by only two dynamic variables.  One variable can be described as the cilium's phase as it provides a map to a particular cilium shape in a prescribed cyclic sequence.  In this study, we use the Fulford and Blake cilium beat \cite{Fulford1986}, building from previous simulations \cite{Ito2019,Omori2020} where the cilium kinematics are prescribed.  The other variable describes the overall orientation of the sequence relative to the underlying surface, allowing for the inclusion of elasticity through a torsional spring.  By varying this spring stiffness, our simulations reveal that below a critical spring stiffness, there is bistability between an azimuthally propagating diaplectic wave, and a polar propagating symplectic-like wave.  Above the critical stiffness, only the diaplectic wave is found to emerge, though multiple wave numbers are possible.  Examining the hydrodynamic performance of these states demonstrates that the diaplectic wave provides higher hydrodynamic efficiency, which we link to its shorter wavelength rather than the specific direction of propagation.  Additionally, we explore the orientation of the cilium beat plane relative to the ciliate body axis -- a feature linked to the rotation of \textit{Volvox} \cite{goldstein2015}, and show that it does not affect the overall emergent coordination and has a minimal impact on the swimming speed for symplectic coordination.   Thus, orienting the beat plane relative to the body axis provides an effective route to introducing swimmer rotation, an important ingredient in \textit{Volvox} phototaxis \cite{drescher2010}, without sacrificing overall swimming speed.

\section{Model}
To begin, we describe the construction of the model ciliate, starting with the formulation of the filament oscillator model and ending with a validation of our numerical implementation through comparison with results from the literature.  

\subsection{Filament Oscillator Model}



\begin{figure}[h!]
    \centering
    \includegraphics[width=\textwidth]{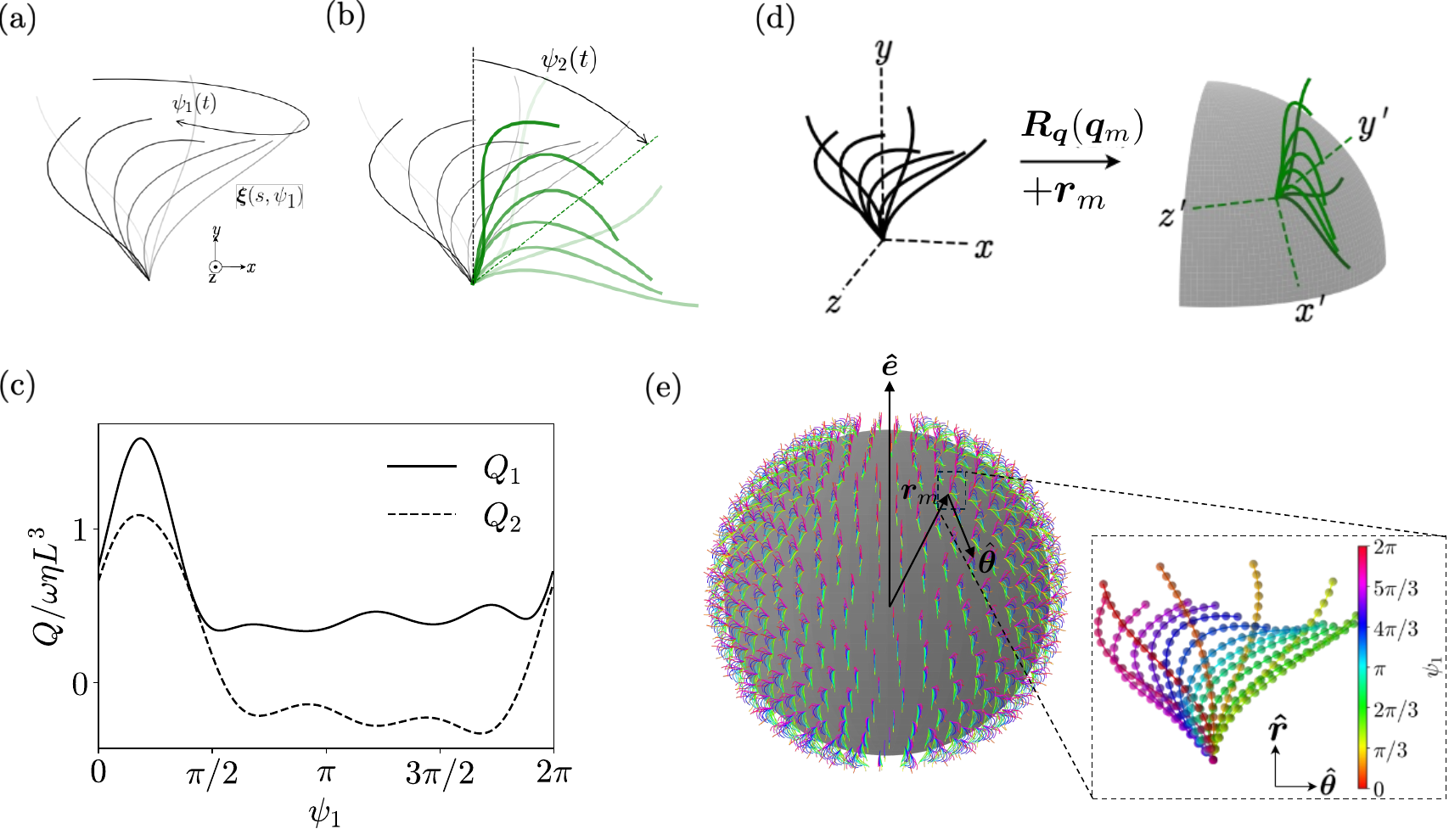}
    \caption{(a) Cilium shapes given by the Fulford and Blake beat for different values of the phase, $\psi_1$.  (b)  The angle $\psi_2$ rotates the entire sequence about an axis perpendicular to the beat plane.  (c)  Generalised forces, $Q_1$ and $Q_2$,  used to drive cilium motion as a function of $\psi_1$. (d) To place and orient cilium $m$ on the ciliate surface, the rotation matrix is applied to $\bm{R}_{\bm{q}}(\bm{q}_m)$ the reference beat, where $\bm{q}_m$ is the quaternion describing the rotation for cilium $m$, and the resulting positions are shifted by the vector $\bm{r}_m$.  (e)  An image of the model ciliate showing the distribution of cilia over its surface and the discretisation of each cilium.  }
    \label{fig:fil_illustration}
\end{figure}

We begin by describing the filament oscillator model for a single cilium with lengthscale\footnote{We note that for the parameterisation from \cite{Fulford1986} that we use in this work, the cilium length is not constant and varies by approximately 5\% during the beat.} $L$ and cross-sectional radius $a$.  The cilium has a planar, time-periodic beat in the $xy$-plane.  Accordingly, the position of the points along the cilium length at time $t$ can be expressed as 
\begin{align}
\bm{\xi}(s, \psi_1(t))= L\left(\xi_1(s, \psi_1(t)) \bm{\hat{x}} + \xi_2(s, \psi_1(t)) \bm{\hat{y}} \right),
\label{eq:FBbeat}
\end{align}
where $s \in [0,1]$ is a parametrisation (not necessarily the arclength) of the cilium centreline.  The quantity $\psi_1(t) \in [0, 2\pi)$ is the cilium's phase and provides the map between time and a particular cilium shape provided by $\bm{\xi}$, as depicted in Fig. \ref{fig:fil_illustration}(a).  We also introduce the angle $\psi_2(t)$, which rotates the cilium about the $z-$axis, see Fig. \ref{fig:fil_illustration}(b).  With this rotation included, the positions along the cilium are
\begin{align}
    \bm{x}(s,\psi_1,\psi_2) = \bm{R}_{\psi}(\psi_2(t))\bm{\xi}(s,\psi_1(t)).\label{eq:filament_pos}
\end{align}
where 
\begin{align}
    \bm{R}_{\psi}(\psi_2) = \begin{bmatrix}
    \cos \psi_2 & -\sin \psi_2 & 0 \\
    \sin \psi_2 & \cos \psi_2 & 0 \\
    0 & 0 & 1
    \end{bmatrix}
\end{align}
is the $z$-axis rotation matrix.  

While the framework we establish is not limited to this choice, in our simulations we utilise the beat provided by Fulford and Blake \cite{Fulford1986} (see again Fig. \ref{fig:fil_illustration}(a)),
\begin{align}
    \xi_i(s, \psi_1) = \sum^{3}_{m=1} \sum^{3}_{n=0}s^m\left( A_{mn}^{(i)} \cos(n\psi_1) + B_{mn}^{(i)} \sin(n\psi_1) \right),
    \label{eq:FBparam}
\end{align}
for $i =1,2$.  We reproduce the coefficients $A_{mn}^{(i)}$ and $B_{mn}^{(i)}$ in Appendix \ref{app:coeffs}.  While this particular cilium beat was tuned to the motion of respiratory tract cilia, we have selected it to connect and compare our results with other recent work on model ciliates \cite{Ito2019, Omori2020}, where this beat was also used.  For the Fulford and Blake beat, the effective stroke occurs for $0 \leq \psi_1 \leq 1.3$ and corresponds to the cilium executing a rapid, nearly rigid rotation.  The effective stroke is followed by recovery for $1.3 \leq \psi_1 \leq 2\pi$ where the cilium exhibits higher curvature as it slowly returns to repeat the effective stroke. %
%

To enable a dynamic simulation in which $\psi_1(t)$ and $\psi_2(t)$ can evolve over time due to interactions between cilia, we must establish equations of motion, which we develop in the spatially discrete setting.  To do so, we first discretise the cilium into $N$ segments whose positions are 
\begin{align}
    \bm{x}_n = \bm{R}_{\psi}(\psi_2)\bm{\xi}(s_n, \psi_1), \label{eq:rotated_shape}
\end{align}for $n =  1\dots N$, or in vector form, $\bm{x} = \left[\bm{x}_1^T \ \bm{x}_2^T\ \dots \ \bm{x}_N^T \right]^T \in \mathbb{R}^{3N \times 1}$.   The velocity of each segment is given by 
\begin{align}
\bm{v}_n = \frac{d\bm{x}_n}{dt} = \bm{k}_1(s_n, \psi_1, \psi_2) \omega_1 + \bm{k}_2(s_n, \psi_1, \psi_2) \omega_2,
\end{align}
where
\begin{align}
\bm{k}_1(s, \psi_1, \psi_2) &= \bm{R}_\psi(\psi_2)\frac{d\bm{\xi}}{d\psi_1}(s,\psi_1) \label{eq:k1} \\
\bm{k}_2(s, \psi_1, \psi_2) &= \frac{d\bm{R}_{\psi}}{d\psi_2}(\psi_2)\bm{\xi}(s,\psi_1), \label{eq:k2}
\end{align}
with $\omega_1 = \dot{\psi}_1$, and $\omega_2 = \dot{\psi}_2$.  Building from these expressions, we can write compactly the velocities of all segments as
\begin{align}
\bm{v} = \bm{K}_1 \omega_1 + \bm{K}_2 \omega_2,  \label{eq:v=k1w1+k2w2}
\end{align}
where $\bm{v} = \left[\bm{v}_1^T \ \bm{v}_2^T\ \dots \ \bm{v}_N^T \right]^T \in \mathbb{R}^{3N \times 1}$, $\bm{K}_1 = \left[\bm{k}^T_1(s_1, \psi_1, \psi_2) \ \bm{k}^T_1(s_2, \psi_1, \psi_2) \ \dots \ \bm{k}^T_1(s_N, \psi_1, \psi_2)\right]^T\in \mathbb{R}^{3N \times 1}$, and similarly, $\bm{K}_2 = \left[\bm{k}^T_2(s_1, \psi_1, \psi_2) \ \bm{k}^T_2(s_2, \psi_1, \psi_2) \ \dots \ \bm{k}^T_2(s_N, \psi_1, \psi_2)\right]^T\in \mathbb{R}^{3N \times 1}$.  

The equations of motion for $\psi_1(t)$ and $\psi_2(t)$ can be derived using an approach closely related to the Lagrangian mechanics of active systems \cite{Solovev2021}. Using the typical values of cilium length and beat frequency from \cite{brumley2012hydrodynamic}, $L \sim 20 \times 10^{-6}\ \textrm{m}$ and $\omega_0 = 2\pi f_0  \sim 66\pi\ \textrm{s}^{-1}$~, 
along with the kinematic viscosity $\nu \sim 1 \times 10^{-6}\ \textrm{m}^2/\textrm{s}$, the Reynolds number associated with cilium motion is $Re = L^2 f_0/\nu \approx 10^{-2}$. As a result, we may take the fluid flow to be described by the Stokes equations.  At each instant in time, there is a linear, but configuration dependent, relations between segment forces, $\bm{\lambda}_n$ for $n = 1, \dots, N$, and their velocities such that
\begin{align}
    \bm{M} \bm{\lambda} = \bm{v} \label{eq:segMob}
\end{align}
where $\bm{M}(\bm{x})$ is the $3N \times 3N$ symmetric mobility matrix for the cilium segments and $\bm{\lambda} = \left[\bm{\lambda}_1^T \ \bm{\lambda}_2^T\ \dots \ \bm{\lambda}_N^T \right]^T \in \mathbb{R}^{3N \times 1}$.  Assuming that the generalised forces $Q_1 = \bm{K}^T_1 \bm{\lambda}$ and $Q_2=\bm{K}^T_2 \bm{\lambda}$ are known, we see that from \eqref{eq:v=k1w1+k2w2} and \eqref{eq:segMob}, cilium motion at each instant in time will be given by the saddle point system,
\begin{align}
\begin{bmatrix}
\bm{M} & -\bm{K}_1  & -\bm{K}_2 \\
-\bm{K}^T_1 & 0 & 0 \\
-\bm{K}^T_2 & 0 & 0 
\end{bmatrix}
\begin{bmatrix}
\bm{\lambda}  \\
\omega_1 \\
\omega_2
\end{bmatrix} =
\begin{bmatrix}
\bm{0}  \\
-Q_1 \\
-Q_2
\end{bmatrix}.
\label{eq:singcilsaddle}
\end{align}
From $\omega_1$ and $\omega_2$, $\psi_1$ and $\psi_2$ can be advanced by integrating $d\psi_i/dt = \omega_i, i=1,2$.  

What remains is to determine the generalised forces, $Q_1(\psi_1)$ and $Q_2(\psi_1)$, required to drive the cilium through its beat at the desired rate.  Specifically, we solve for $Q_1$ and $Q_2$ by rearranging \eqref{eq:singcilsaddle} to establish a resistance problem for a single cilium with $\psi_2 = 0$, $\omega_2(t) = 0$, and $\omega_1(t) = \omega_0$, where $\omega_0$ is a positive constant.  We perform this computation at $1000$ equispaced values of $\psi_1$ for $0 \leq \psi_1 < 2\pi$ to tabulate values of $Q_1$ and $Q_2$ that we use to generate interpolants in ciliate simulations.   We note that even though $\omega_0$ is a constant, the cilium velocity is not constant due to the $\psi_1$ dependence of $\bm{K}_1$ and $\bm{K}_2$.  We incorporate the effect of a nearby surface in the computation of generalised forces by evaluating $\bm{M}$ using the pairwise, wall-corrected RPY mobility matrix \cite{Swan2007}, such that
\begin{align}
\bm{v}_n = \sum_{m = 1}^N \bm{M}^{\mathrm{RPY-wall}}_{nm}\bm{\lambda}_{m}.
\end{align}
The resulting values of $Q_1(\psi_1)$ and $Q_2(\psi_1)$ are shown in Fig. \ref{fig:fil_illustration}(c).  We see that both $Q_1$ and $Q_2$ attain their maximum values during the effective stroke due to the rapid movement of the cilium during this portion of the beat cycle.  During recovery, $Q_1$ is lower but still positive as $\psi_1$ is increasing with time, while $Q_2$ has changed sign, indicating an opposite force is required to keep $\psi_2$ fixed when the cilium moves in the opposite direction.  

Lastly, in order to limit changes in $\psi_2$ as well as incorporate a notion of elasticity in the ciliate simulations, we include in the model the generalised force associated with a linear torsional spring that returns $\psi_2$ to zero.  Specifically, we take $Q_k(\psi_2) = -k_{\psi}\psi_2$, where $k_{\psi}$ is the spring constant, and add it to $Q_2$.

\subsection{Model ciliate dynamics}
Having established the filament oscillator model, we now employ it to compute the dynamics of a model spherical ciliate whose motion is driven by many filament oscillators.  We begin by describing the motion of the rigid surface to which the cilia are attached.  The ciliate is taken to have centre position $\bm{Y}(t)$, body axis $\bm{\hat{e}}(t)$, and radius $R$.  The orientation of the ciliate is given by the quaternion $\bm{q}(t)=\left[ q_0, q_1, q_2, q_3 \right]^T$.  

Following the rigid multiblob method \cite{BalboaUsabiaga2016,delmotte2025modeling}, we discretise the spherical ciliate surface into $P$ elements, where the position of element $p$ is given by
\begin{align}
    \bm{y}_p(t) = \bm{Y}(t) + \bm{R}_{\bm{q}} (\bm{q}(t)) \bm{r}_p \label{eq:y=Y+Rr}
\end{align}
where $\bm{r}_p$ is the position of element $p$ relative to $\bm{Y}$ at $t=0$. In Appendix \ref{sec:kmeans}, we describe the $k$-means-based algorithm that we use to distribute $\bm{r}_p$ on the sphere.  The rotation matrix $\bm{R}_{\bm{q}}(\bm{q})$ is related to the quaternion through
\begin{align}
    \bm{R}_{\bm{q}}(\bm{q}) = \begin{bmatrix}
        1-2q_2^2-2q_3^2 & 2(q_1q_2-q_3q_0) & 2(q_1q_3+q_2q_0) \\
        2(q_1q_2+q_3q_0) & 1-2q_1^2-2q_3^2 & 2(q_3q_2-q_1q_0) \\
        2(q_1q_3-q_2q_0) & 2(q_3q_2+q_1q_0) & 1-2q_2^2-2q_1^2)
    \end{bmatrix}
\end{align}
In accordance with rigid body dynamics, the velocity of element $p$ is
\begin{align}
    \bm{u}_p = \bm{V}(t) + \bm{\Omega}(t) \times \left(\bm{y}_p(t) - \bm{Y}(t)\right)
\end{align}
where $\bm{V}(t)$ is the translational velocity of the ciliate, and $\bm{\Omega}(t)$ is its angular velocity.  We may express this compactly for all $P$ elements as
\begin{align}
    \bm{u} = \bm{K}_S(\bm{y})\bm{U}
\end{align}
where $\bm{u} = [\bm{u}^T_1 \ \bm{u}^T_2 \ \dots \ \bm{u}^T_P]^T \in \mathbb{R}^{3P \times 1}$,  $\bm{y} = [\bm{y}^T_1 \ \bm{y}^T_2 \ \dots \ \bm{y}^T_P]^T\in \mathbb{R}^{3P \times 1}$, $\bm{U} = [\bm{V}^{T} \ \ \bm{\Omega}^T]^T\in \mathbb{R}^{6 \times 1}$.  The $3P \times 6$ matrix $\bm{K}_S$ provides the linear relationship between the rigid body motion of the ciliate and the velocity of each surface element \cite{BalboaUsabiaga2016,delmotte2025modeling}. 

To complete the ciliate construction, we distribute $M$ cilia on the surface of the sphere such that the effective stroke of each is directed toward the posterior pole of the ciliate.  The base of cilium $m$ has position $\bm{r}_m$ relative to $\bm{Y}$, which we assign using the algorithm described in Appendix \ref{sec:kmeans}.  The orientation of its beat in the ciliate body axes is given by a constant rotation described by the quaternion $\bm{q}_m$.  As depicted in Figs. \ref{fig:fil_illustration}(d) and \ref{fig:fil_illustration}(e), using $\bm{r}_m$ and $\bm{q}_m$, we may express the position of segment $n$ on cilium $m$ as
\begin{align}
    \bm{x}_{nm} = \bm{Y} + \bm{R}_{\bm{q}}(\bm{q})\left(\bm{r}_m +  \bm{R}_{\bm{q}}(\bm{q}_m)\bm{x}_n\left(\psi^{(m)}_1, \psi^{(m)}_2\right)\right). 
\end{align}
where $\psi^{(m)}_1$ and $\psi^{(m)}_2$ are the phase and orientation variables, respectively, for cilium $m$, and $\bm{x}_n$ is given by \eqref{eq:rotated_shape}.  The velocity of segment $n$ on cilium $m$ is therefore,
\begin{align}
    \bm{v}_{nm} = \bm{V} + \bm{\Omega} \times \left(\bm{r}_m +  \bm{R}_{\bm{q}}(\bm{q}_m)\bm{x}_n\left(\psi^{(m)}_1, \psi^{(m)}_2\right)\right) + \bm{R}_{\bm{q}}(\bm{q})\bm{R}_{\bm{q}}(\bm{q}_m)\bm{v}_n\left(\psi^{(m)}_1, \psi^{(m)}_2\right).
\end{align}
We combine the velocities of all $NM$ segments into a $3NM \times 1$ vector, $\bm{v}$, and, using \eqref{eq:v=k1w1+k2w2}, express $\bm{v}$ as 
\begin{align}
    \bm{v} = \bm{K}_C \bm{U} + \bm{\widetilde{K}}_1 \bm{\omega}_1 + \bm{\widetilde{K}}_2 \bm{\omega}_2 
\end{align}
where $\bm{v} = [\bm{v}^T_{11} \ \bm{v}^T_{21} \ \dots \ \bm{v}^T_{NM}]^T \in \mathbb{R}^{3NM \times 1}$, $\bm{\omega}_1 = [\omega^{(1)}_{1} \ \omega^{(2)}_{1}\ \dots \ \omega^{(M)}_{1}]^T \in \mathbb{R}^{M \times 1}$, and $\bm{\omega}_2 = [\omega^{(1)}_{2} \ \omega^{(2)}_{2}\ \dots \ \omega^{(M)}_{2}]^T\in \mathbb{R}^{M \times 1}$.  The rectangular matrices $\bm{K}_{C} \in \mathbb{R}^{3MN \times 6}$ , $\bm{\widetilde{K}}_1 \in \mathbb{R}^{3MN \times M}$ and $\bm{\widetilde{K}}_2 \in \mathbb{R}^{3MN \times M}$ map the ciliate rigid body motion, $\bm{\omega}_1$, and $\bm{\omega}_2$, respectively, to the segment velocities.  The matrix $\bm{K}_{C}$ fulfills the same role as $\bm{K}_{S}$ does for the surface elements, while the non-zero entries of the $m$th columns of $\bm{\widetilde{K}}_1$ and $\bm{\widetilde{K}}_2$ are given by $\bm{K}^{(m)}_1$ and $\bm{K}^{(m)}_2$, respectively, the vectors $\bm{K}_1$ and $\bm{K}_2$ corresponding to the $m$th cilium.  The expressions for these matrices are provided in the Appendix \ref{appendix:geometric_matrices}.  


With expressions for the surface element and segment velocities established, we can formulate a mobility problem to compute $\bm{U}$, $\bm{\omega}_1$, and $\bm{\omega}_2$ that accounts for the hydrodynamic interactions between all cilia and the ciliate surface.  Again, since fluid inertia is negligible, there will be a linear relationship, 
\begin{align}
\begin{bmatrix}
\bm{u} \\
\bm{v}
\end{bmatrix}
= \begin{bmatrix}
\bm{M}_{SS} & \bm{M}_{SC}\\
\bm{M}_{CS} & \bm{M}_{CC}
\end{bmatrix}
\begin{bmatrix}
\bm{\nu} \\
\bm{\lambda}
\end{bmatrix},
\label{eq:mobcilsurf}
\end{align}
between the segment and surface element velocities, $\bm{u}$ and $\bm{v}$, respectively, and the segment and surface element forces, $\bm{\lambda}$ and $\bm{\nu}$, respectively, where now $\bm{\lambda} \in \mathbb{R}^{3NM \times 1}$ and $\bm{\nu} \in \mathbb{R}^{3P \times 1}$.  The action of the configuration-dependent mobility matrices, $\bm{M}_{SS} \in \mathbb{R}^{3P\times 3P}$, $\bm{M}_{SC}\in \mathbb{R}^{3P\times 3NM}$, $\bm{M}_{CS}\in \mathbb{R}^{3NM\times 3P}$ and $\bm{M}_{CC}\in \mathbb{R}^{3NM \times 3NM}$ providing the surface-surface, surface-cilia, cilia-surface, and cilia-cilia hydrodynamic interactions, respectively, is computed using the force-coupling method \cite{maxey2001,Su2024} as described in the next section.  The total force and torque on the ciliate are related to $\bm{\lambda}$ and $\bm{\nu}$ through
\begin{align}
\bm{F} = \bm{K}_{S}^T \bm{\nu} + \bm{K}_{C}^T \bm{\lambda},
\label{eq:totalFcil}
\end{align}
where $\bm{F} \in \mathbb{R}^{6 \times 1}$, while the generalised forces on all cilia segments,
\begin{align}
\bm{Q}_1 &= [Q_1(\psi_1^{(1)}) \ Q_1(\psi_1^{(2)}) \ \dots \ Q_1(\psi_1^{(M)})]^T \in \mathbb{R}^{M \times 1}, \\
\bm{Q}_2 &= [Q_2(\psi_1^{(1)}) \ Q_2(\psi_1^{(2)}) \ \dots \ Q_2(\psi_1^{(M)})]^T \in \mathbb{R}^{M \times 1},
\end{align}
and $\bm{Q}_k = -k_{\psi}[\psi_2^{(1)} \ \psi_2^{(2)} \ \dots \ \psi_2^{(M)}]^T \in \mathbb{R}^{M \times 1}$, are related to $\bm{\lambda}$ through
\begin{align}
\bm{Q}_1 &= \bm{\widetilde{K}}_{1}^T \bm{\lambda} \\
\bm{Q}_2 +\bm{Q}_k  &= \bm{\widetilde{K}}_{2}^T \bm{\lambda} 
\label{eq:Qcil}
\end{align}
Using \eqref{eq:mobcilsurf}, \eqref{eq:totalFcil}, and \eqref{eq:Qcil}, we are in a position to formulate a mobility problem to determine cilia and ciliate motion at each instant.  Specifically, from $\bm{Y}$, $\bm{q}$, and $\psi_1^{(m)}$ and $\psi_2^{(m)}$ for $m = 1, \dots, M$, we first compute $\bm{Q}_1$, $\bm{Q}_2$ and $\bm{Q}_k$, as well as the entries of the matrices $\bm{K}_S$, $\bm{K}_C$, $\bm{\widetilde{K}}_1$ and $\bm{\widetilde{K}}_2$.  For the case where the ciliate is free to swim, we insist that the ciliate is force- and torque-free and set $\bm{F} = 0$.  The resulting motion of the model ciliate is found by solving the saddle-point system
\begin{align}
\begin{bmatrix} \bm{M}_{SS} & \bm{M}_{SC} & -\bm{K}_S & \bm{0} & \bm{0} \\
\bm{M}_{CS} & \bm{M}_{CC} & -\bm{K}_C & -\bm{\widetilde{K}}_1 & -\bm{\widetilde{K}_2} \\
-\bm{K}_{S}^T & -\bm{K}_{C}^T & \bm{0} & \bm{0} & \bm{0} \\
\bm{0} & -\bm{\widetilde{K}_1}^T & \bm{0} & \bm{0} & \bm{0} \\
\bm{0} & -\bm{\widetilde{K}_2}^T & \bm{0} & \bm{0} & \bm{0} \\
\end{bmatrix} \begin{bmatrix} \bm{\nu} \\
\bm{\lambda} \\
\bm{U} \\
\bm{\omega}_1 \\
\bm{\omega}_2
\end{bmatrix}
= \begin{bmatrix}\bm{0} \\
\bm{0} \\
-\bm{F} \\
-\bm{Q}_1 \\
-\left(\bm{Q}_2 + \bm{Q}_k\right)
\end{bmatrix},\label{eq:full_saddle_point}
\end{align}
and integrating the differential equations
\begin{align}
\begin{split}
\frac{d\bm{Y}}{dt} &= \bm{V}, \\
\frac{d\bm{q}}{dt} &= \frac{1}{2}(0,\bm{\Omega})\bullet \bm{q} \\
\frac{d\bm{\psi}_1}{dt} &= \bm{\omega}_1,\\
\frac{d\bm{\psi}_2}{dt} &= \bm{\omega}_2,
\end{split} \label{eq:time_derivatives}
\end{align}
to advance the position and orientation of the ciliate, as well as the phase and orientation of each oscillator.  For the cases where the ciliate is held fixed, we instead take $\bm{U} = 0$ and rearrange the saddle point system accordingly.

\subsection{Numerical methods}
To implement the ciliate model, we rely on several numerical algorithms to apply the mobility matrices, solve the saddle-point system, and integrate the differential equations \eqref{eq:time_derivatives}.  We summarise these methods here and refer the reader to other publications for more details.  

\subsubsection{Solving the saddle-point system}
At the heart of the computation of ciliate motion is the linear saddle-point system in \eqref{eq:full_saddle_point}.  Appearing in this expression is the $3(MN + P) \times 3(MN + P)$ mobility matrix, 
\begin{align}
\bm{M} = \begin{bmatrix} \bm{M}_{SS} & \bm{M}_{SC} \\
\bm{M}_{CS} & \bm{M}_{CC}
\end{bmatrix}
\end{align}
that provides the hydrodynamic interactions between all cilia segments and surface elements.  In our computations, the action of the mobility matrix of the vectors of $\bm{\lambda}$ and $\bm{\nu}$ is provided by the force-coupling method (FCM) \cite{maxey2001,Su2024}.  In FCM, each segment or element is represented as a Gaussian distribution,
\begin{align}
\Delta(\mathbf{x}) = (2 \pi \sigma^2)^{-3/2} \exp \left( -\lVert \mathbf{x} \rVert^2/(2 \sigma ^2)\right), 
\end{align}
of force in the Stokes equations with $\sigma = a / \sqrt{\pi}$, such that the resulting fluid flow at a point $\mathbf{x}$ in the fluid domain, $\Omega$, is given by,
\begin{align}
-\bm{\nabla} p + \eta \nabla^2 \mathbf{u} + \sum_{m=1}^M\sum_{n=1}^N \Delta(\mathbf{x} - \bm{x}_{mn})\bm{\lambda}_n + \sum^P_{p=1} \Delta(\mathbf{x} - \bm{y}_p)\bm{\nu}_p &= 0\\
\bm{\nabla} \cdot \mathbf{u} &= 0.
\end{align}
The fluid flow is then volume averaged against the same Gaussian distributions to obtain the velocities of the segments and surface elements,
\begin{align}
\bm{v}_{nm} &= \int_\Omega \mathbf{u}(\mathbf{x})\Delta(\mathbf{x} - \bm{x}_{mn})d^3\mathbf{x} \\
\bm{u}_{p} &= \int_\Omega \mathbf{u}(\mathbf{x})\Delta(\mathbf{x} - \bm{y}_p)d^3\mathbf{x}. 
\end{align}
To enable efficient computation, the application of $\bm{M}$ through FCM is performed in a triply-periodic domain of side length $H$ to take advantage of the fast FCM algorithm \cite{Su2024} and the hydrodynamic radius of the surface elements and cilium segments are taken to be $a$.

Since fast FCM is a matrix-free method, it is convenient to use the Krylov subspace method GMRES to solve the saddle point system \eqref{eq:full_saddle_point}.  We precondition the system using a right-preconditioner that is constructed by replacing $\bm{M}$ in \eqref{eq:full_saddle_point} with a diagonal mobility matrix, $\frac{1}{6\pi a\eta} \bm{I} \in \mathbb{R}^{3(MN + P) \times 3(MN + P)}$.  Based on this diagonal mobility matrix, we are able to obtain explicit expression for the inverse of the preconditioner, which we can then use at each GMRES iteration. 

\subsubsection{Time integration}

After solving the saddle point system, the differential equations \eqref{eq:time_derivatives} can be integrated to update the ciliate position and orientation, as well as the phases for all cilia.  Due to numerical stiffness that can arise for large values of $k_{\psi}$, we utilise an implicit scheme.  Specifically, we use the implicit second-order BDF scheme for the position and phases such that at step $k$, we have 
\begin{align}
\bm{Y}^k  - \frac{4}{3} \bm{Y}^{k-1} + \frac{1}{3}\bm{Y}^{k-2} &= \frac{2}{3}\Delta t \bm {V}^{k} \\
\bm{\psi}_1^k  - \frac{4}{3} \bm{\psi}_1^{k-1} + \frac{1}{3}\bm{\psi}_1^{k-2} &= \frac{2}{3}\Delta t \bm {\omega}_1^{k} \\
\bm{\psi}_2^k  - \frac{4}{3} \bm{\psi}_2^{k-1} + \frac{1}{3}\bm{\psi}_2^{k-2} &= \frac{2}{3}\Delta t \bm {\omega}_2^{k} \\
\end{align}
and a geometric second-order BDF (see \cite{schoeller2021methods}) for the unit quaternion describing ciliate orientation.  Here, we relate $\bm{q}^k$ to $\bm{q}^{k-1}$ using the Lie algebra element, $\bm{\theta}$, such that
\begin{align}
\bm{q}^{k} = \exp(\bm{\theta}^k) \bullet \bm{q}^{k-1},
\end{align}
where the exponential map is given by, 
\begin{align}
\exp(\bm{\theta}) = \left( \cos \left(\frac{\lVert \bm{\theta}\rVert}{2} \right), \sin \left(\frac{\lVert \bm{\theta}\rVert}{2}  \right) \frac{\bm{\theta}}{\lVert \bm{\theta}\rVert}\right).
\end{align}
The Lie algebra element is updated through,
\begin{align}
\bm{\theta}^{k} - \frac{1}{3}\bm{\theta}^{k-1} = \frac{2}{3} \textrm{dexp}^{-1}_{\bm{\theta}^{k}}(\bm{\Omega}^k),
\end{align}
and the differential of the inverse of the exponential map is,
\begin{align}
\textrm{dexp}^{-1}_{\bm{\theta}}(\bm{\Omega}) = \bm{\Omega} - \frac{1}{2}\bm{\theta}\times \bm{\Omega} - \frac{1}{2\lVert \bm{\theta}\rVert^2}\left(\lVert \bm{\theta}\rVert \cot\left(\frac{\lVert \bm{\theta}\rVert}{2} \right) - 2\right)\bm{\theta} \times \left( \bm{\theta} \times \bm{\Omega}\right).
\end{align}
The update equations establish a system of nonlinear equations whose solution provides $\bm{Y}^k, \bm{\psi}_1^k, \bm{\psi}_2^k$, and $\bm{\theta}^k$ (and hence $\bm{q}^k$).  We solve this system of equations iteratively using Broyden's method.

\subsection{Validation}

 Before running the full simulations of the model ciliate, we perform several numerical tests, comparing with known analytical solutions, as well as previous simulation results for ciliates \cite{Omori2020}.  

\subsubsection{Settling sphere}

We first test the resolution of the no-slip boundary condition by computing the velocity of a rigid sphere subject to an applied force.  We consider the saddle point system \eqref{eq:full_saddle_point} for a sphere of radius $R$ discretised by $P$ surface elements, each with hydrodynamic radius $a$, in the absence of any attached filaments, i.e. $M = 0$.  The sphere is subject to a given applied force $\bm{F} = F_0 \hat{\bm{z}}$ and to completely remove the effects of periodicity, the calculation is performed by evaluating pairwise the expressions for the FCM mobility matrix (see \cite{Su2024}) for an unbounded domain.   Further, as we increase $P$, we preserve the ratio $R/a = 1.831\sqrt{P}$ so as to systematically decrease the hydrodynamic radius of the surface elements as they increase in number.  Fig. \ref{fig:validation}(a) shows the difference in the translational speed, $\bm{V} \cdot \hat{\bm{z}}$, and the Stokes settling speed, $W = F_0/(6\pi\eta R)$, as a function of $P$.  We see that a relative error of less than 1\% is achieved with $P = 10000$ and the error decreases as $\sim P^{-1/2}$, consistent with the rigid multiblob \cite{BalboaUsabiaga2016,delmotte2025modeling} and immersed boundary \cite{mori2008convergence} methods.

\subsubsection{Steady spherical squirmer}

We also perform a test in which, instead of applying a force to the sphere, we prescribe the axisymmetric, tangential surface velocity, $\bm{u}_{\theta}(\theta) = B_1 \sin \theta \bm{\hat {\theta}}$, where $\bm{\hat {\theta}}$ is the unit vector in the polar direction and $\theta$ is the polar angle.  This surface velocity corresponds to a neutral steady squirmer \cite{Blake1971}.  We again solve the saddle-point system to find the rigid body motion of the sphere with $\bm{F} = \bm{0}$ and compare the resulting translational velocity to the known analytical value, $V=2B_1/3$, as $P$ increases with $R/a = 1.831\sqrt{P}$.  The error in the swimming speed is shown in Fig. \ref{fig:validation}(b).  We again observe the correct rate of convergence and errors of less than 1\% for $P \geq 20000$.

\subsubsection{Model ciliated sphere}

\begin{figure}[h!]
    \centering
    \includegraphics[width=\textwidth]{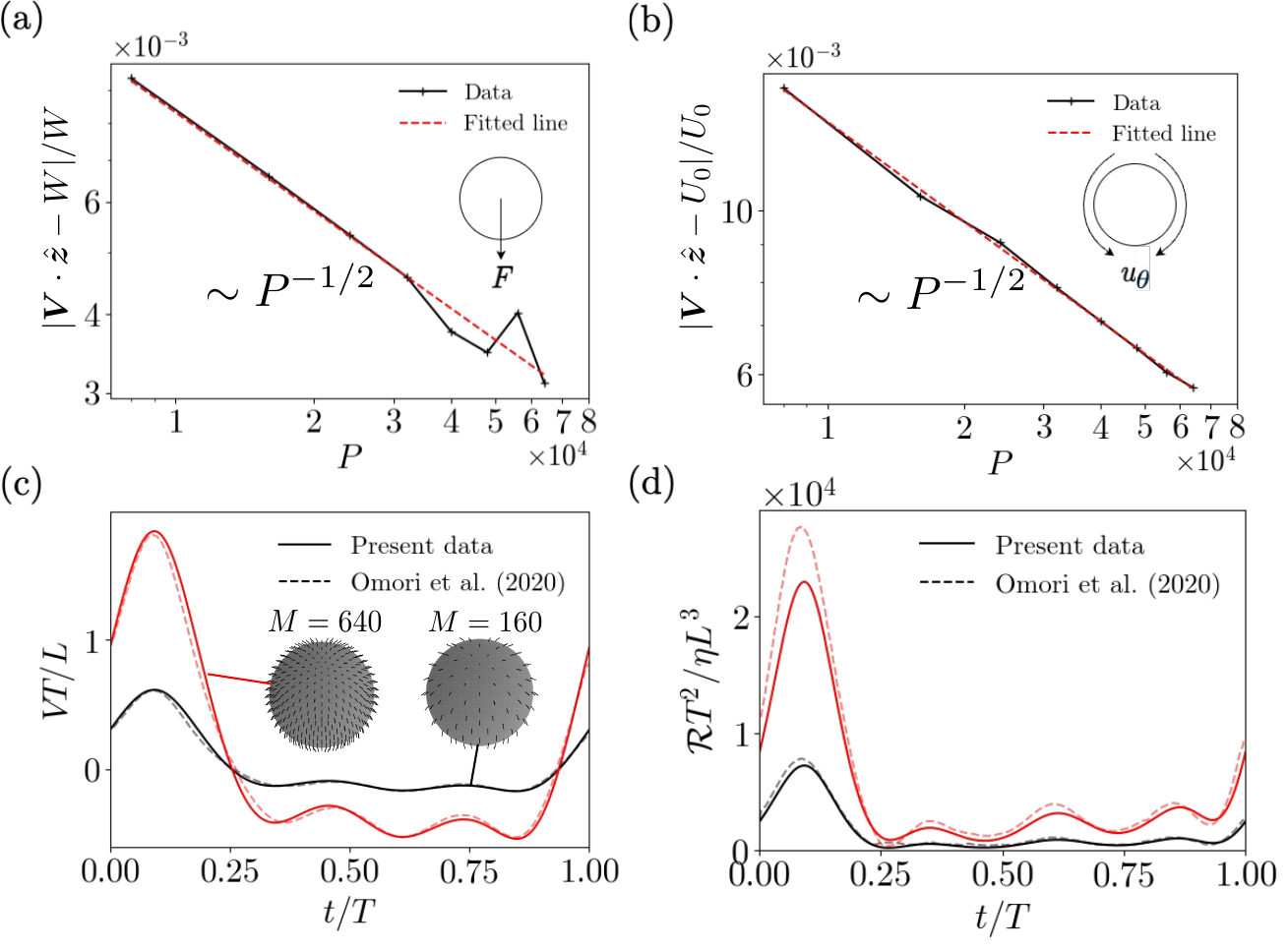}
    \caption{(a)\label{fig:validation_settling} The relative error in the settling speed of a rigid sphere as a function of the number of surface elements. The error decays like $\sim P^{-1/2}$.  (b)\label{fig:validation_squirmer} The relative error in the swimming speed for a neutral squirmer as a function of $P$.  Again, the error is found to be $\sim P^{-1/2}$.   A direct comparison with \cite{Omori2020} of the (c) swimming speed and (d) viscous dissipation for model ciliates with $M=160$ and $M=640$ cilia beating in synchrony}
    \label{fig:validation}
\end{figure}

As a final test of our methodology, we compute the motion of a spherical ciliate propelled by Fulford--Blake cilia and compare with previous results from \cite{Ito2019,Omori2020}. We replicate their ciliate geometry by setting $R/L=10.0$ and utilise the same cilia placement on the spherical surface.  We consider cases where $M=160$ and $640$ with the effective strokes oriented toward the posterior of the ciliate.  In our computations, each cilium is discretised into $N=40$ segments, and to avoid cilia segments overlapping during the beat, we set $\Delta L = 2.6a$. This results in a slenderness $a/L = 1/101.4$, which is comparable to the value of $a/L = 1/100$ used in \cite{Ito2019,Omori2020}. The spherical surface is discretised using $P=40962$ elements.  Cilia motion is prescribed such that all cilia are synchronised, i.e. $\omega_{1}^{(m)}= \omega_0$ with $\psi_{1}^{(m)}(0) = 0$ and $\psi_{2}^{(m)}(t) = 0$ for all $m$, and the resulting ciliate velocity, $\bm{V}(t)$, angular velocity, $\bm{\Omega}(t)$, and viscous dissipation,
\begin{align}
\mathcal{R}(t) = (\bm{\lambda}(t))^T\bm{v}(t) + (\bm{\nu}(t))^T\bm{u}(t),
\end{align}
for one period, $T  = 2\pi/\omega_0$ are computed under the condition $\bm{F} = 0$.  The dimensionless values of the velocity in the body-axis direction $V(t) = \bm{V}(t) \cdot \bm{\hat e}(t)$ and $\mathcal{R}(t)$ over one period are shown in Figs.\ref{fig:validation}(c) and \ref{fig:validation}(d), respectively.  We see that the resulting swimming speed during the beat cycle closely matches the results from \cite{Omori2020}, showing a peak speed in the $\bm{\hat{e}}-$direction during the effective stroke, and motion in the opposite direction when all cilia are in recovery.  The viscous dissipation values also match those from \cite{Omori2020}, however, there is more discrepancy here than for the swimming speed, especially at the peak of the effective stroke.  We suspect that the difference in viscous dissipation can be attributed to the differences in hydrodynamic model, as well as the different levels of discretisation that are used.  We replicated the original filament seedings precisely, derived by iteratively dividing the edges and faces of an icosahedron and explored increasing the resolution by adding more surface blobs.  We also adjusted the distance of the ciliary bases (i.e. the first segment) from the surface.  None of these geometric changes improved the comparison.  We also note that due to the $L^3$ scaling that appears in $\mathcal{R}$, small differences in length can greatly affect the reported nondimensionalised viscous dissipation.


\section{Cilia coordination}

In this section, we perform fully dynamic simulations to study cilia coordination on the model ciliate.  We consider simulations where we fix the ciliate geometry and examine coordination as we vary the dimensionless stiffness parameter, 
\begin{align}
k = k_{\psi}/(\eta \omega_0 L^3),
\end{align}
in the range $k \in[0.005, 0.1]$. Using the values of cilium bending rigidity, cilium length and beat frequency 
reported in \cite{Brumley2015}, $K_B = 4 \times 10^{-22}\ \textrm{N}{\cdot}\textrm{m}^2$, 
$L = 20 \times 10^{-6}\ \textrm{m}$, $\omega_0 = 66\pi\ \textrm{s}^{-1}$, respectively, 
as well as the viscosity of water, $\eta = 10^{-3}\ \textrm{N}{\cdot}\textrm{s}/\textrm{m}^2$, 
we estimate the corresponding dimensionless stiffness for cilia as 
$K_B/(\eta \omega_0 L^4) \approx 0.1$. In the simulations, the ciliate has radius $R/L = 7.5$ and is covered with $M = 639$ cilia, numbers comparable to those reported 
for \emph{Volvox aureus}, which has $500-1000$ somatic cells and $R/L = 5 - 15$ \cite{goldstein2015}.  The values of $\bm{q}_m$ for $m = 1,\dots, M$ are set such that the effective strokes of all cilia are toward the posterior pole of the ciliate.  We consider both cases where the ciliate is held fixed ($\bm{U}=\bm{0}$), and where it is allowed to swim freely ($\bm{F}=\bm{0}$).

Informed by the numerical tests presented in the previous section, the ciliate surface is discretised using $P = 40961$ points.  Each cilium is taken to have $N=20$ segments with centre-to-centre segment spacing $2.6a$, so the cilium aspect ratio is $a/L = 1/49.4$.  Finally, the side length, $H$, of the periodic domain is set to $H = 10.8R$.  As a result, the spherical body of the ciliate occupies a volume fraction $4\pi R^3/(3H^3) = 0.0033$, suggesting that the effect of the periodic domain will be limited.


\begin{figure}[h!]
    \centering
    \includegraphics[width=\textwidth]{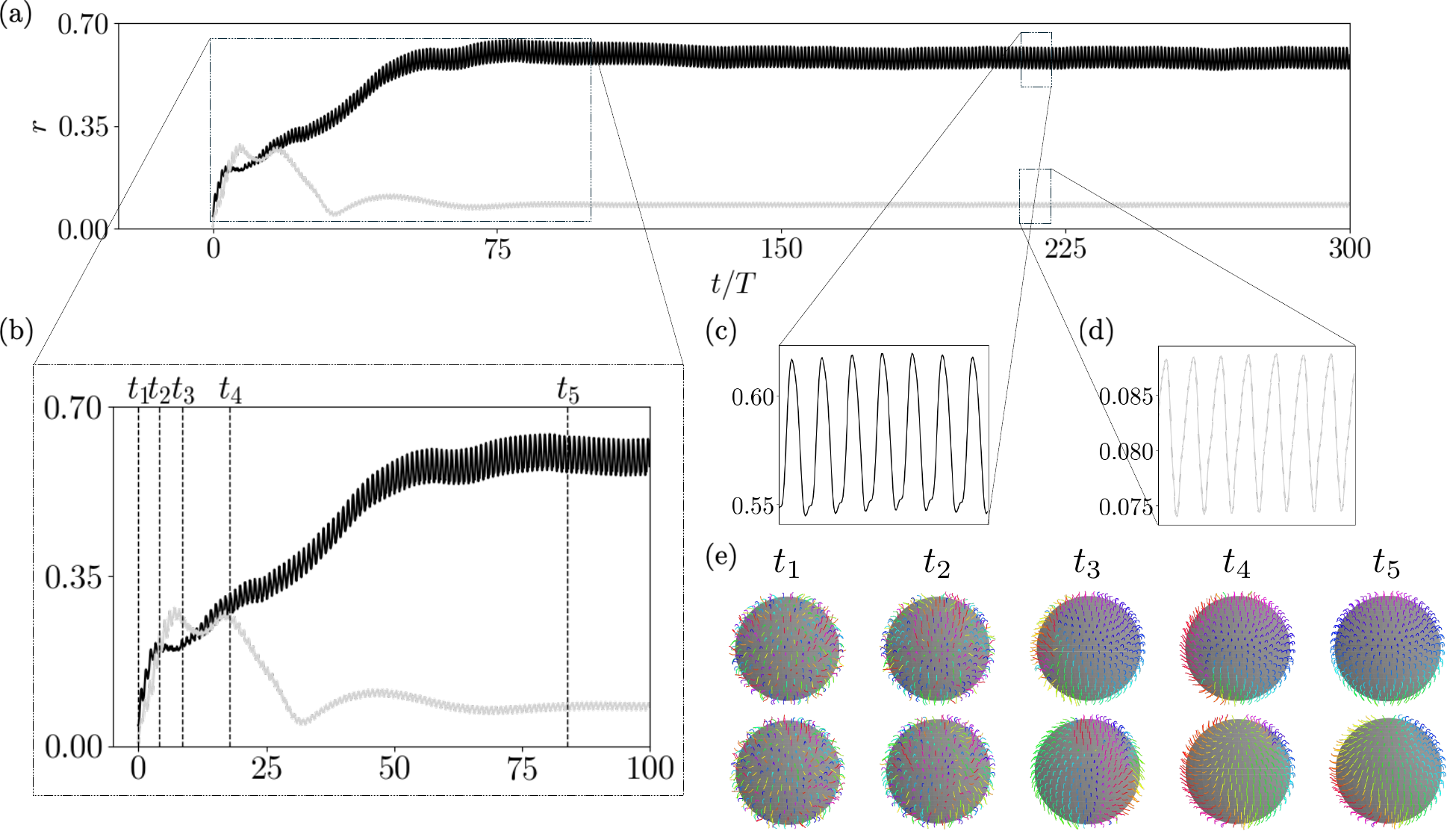}
    \caption{The time evolution of $r$ for two ciliates held fixed with $k=0.005$.  The two cases differ in their initial conditions, leading to the eventual emergence of different states.  (a) \label{fig:roadmap_long} The evolution $r$ over the entire 1000T for each simulation.  The final states persist for several hundred periods. (b) \label{fig:roadmap_zoomin} A zoomed-in view of $r(t)$ for the first 100 periods.  The periodic variations in $r$ for (c)\label{fig:roadmap_symplectic} the symplectic wave and (d) \label{fig:roadmap_diaplectic}the diaplectic wave.  (e) \label{fig:roadmap_phases} The model ciliates at the different times, $t_i$, for $i = 1, \dots, 5$ indicated in panel (a).  Videos showing the emergence can be found in the supplementary materials.}
    \label{fig:roadmap}
\end{figure}

In the simulations, each $\psi_1^{(m)}$ has an initial value drawn randomly from the uniform distribution, $\mathcal{U}(0,2\pi)$, and $\psi_2^{(m)} = 0$ for all $m$.  Simulations are typically run to the final time of $t_f = 1000T$.  To quantify emergent coordination, we compute the Kuramoto order parameter,
    \begin{align}
        r(t) = \left| \frac{1}{M}\sum_{m=1}^M \exp(i\psi_1^{(m)}(t))\right|.
    \end{align}
If the cilia were to have random phases, then $r \approx 0$, while if they are synchronised, then $r = 1$.  Fig. \ref{fig:roadmap}(a) shows the evolution of $r(t)$ for two simulations where the ciliate is held fixed and $k = 0.005$.  We see that for both simulations, after an initial transient period of $100T$, the simulations reach their asymptotically stable states.  Focusing on the initial $100T$ in Fig. \ref{fig:roadmap}(b), we see that the time-evolutions of $r(t)$ for the first 25 periods are qualitatively similar.  After this, however, the $r(t)$ for the simulations diverge from each other and have very different values after approximately 50 periods.  In one case, we see that $r$ oscillates about a value of $r \approx 0.6$ (Fig. \ref{fig:roadmap}(c)), while in the other case we have $r \approx 0.1$ (Fig\ref{fig:roadmap}(d)).  Observing the phases of the cilia in these simulations, see Fig\ref{fig:roadmap}(b), we see that the $r \approx 0.6$ simulation has large patches of nearly synchronised cilia with phases that vary with polar angle.  On the other hand, the simulation with $r \approx 0.1$ has phases that vary with the azimuthal angle.  Given the direction of the effective stroke, we refer to the $r \approx 0.6$ state as a symplectic wave and the $r \approx 0.1$ case as a diaplectic wave.  In both cases, we see that these states persist for hundreds of periods after they emerge (see again Fig. \ref{fig:roadmap}(a)), indicating a bistability between symplectic and diaplectic metachronal waves for $k = 0.005$.  


\begin{figure}[h!]
    \begin{center}
        \includegraphics[width=\textwidth]{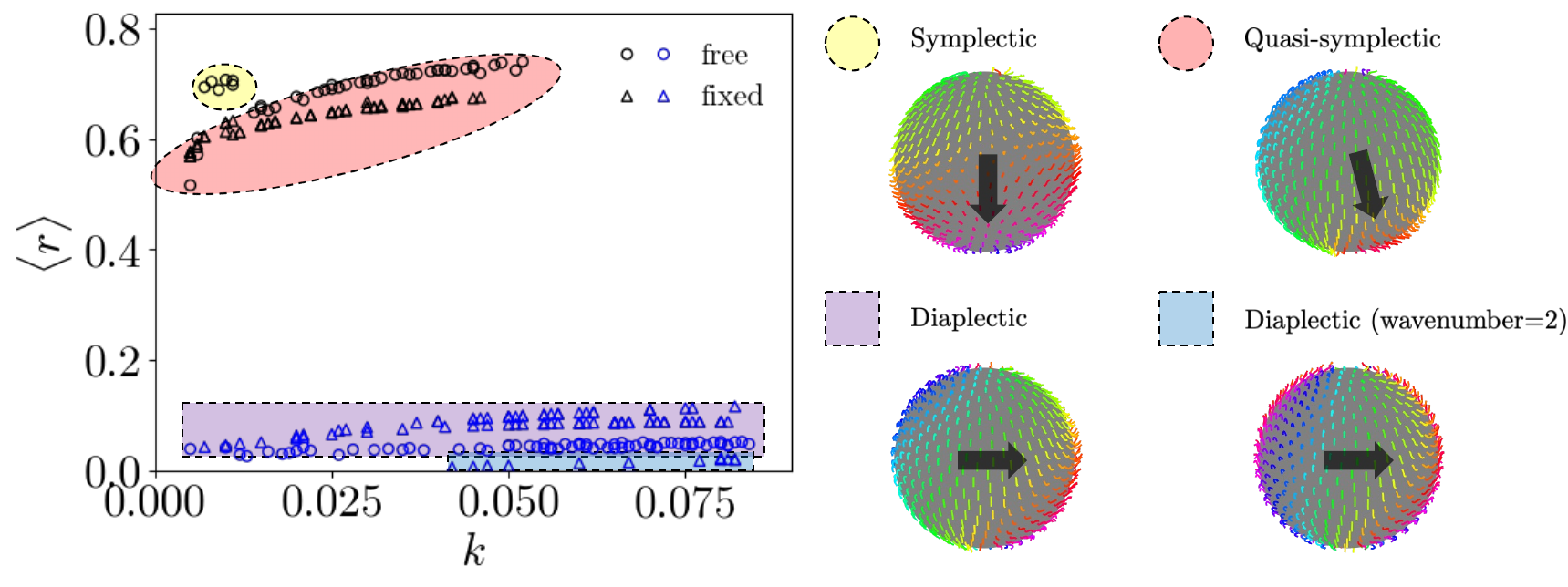}
    \end{center}
    \caption{Time-averaged Kuramoto order parameter, $\langle r \rangle$, as a function of $k$ for the final states of ciliate simulations with $M=639$ cilia and $R/L = 7.5$.  Results are shown for both held fixed and free-to-swim ciliates.  The qualitatively different final states are indicated by the different colours.  Representative cases of these states are also shown.  Notably, below a critical value $k_c \approx 0.055$, both symplectic-like and diaplectic waves are observed.  Above the critical value, only diaplectic waves are observed.}
    \label{fig:order_parameter}
\end{figure}

\begin{figure}[h!]
    \begin{center}
        \includegraphics[width=\textwidth]{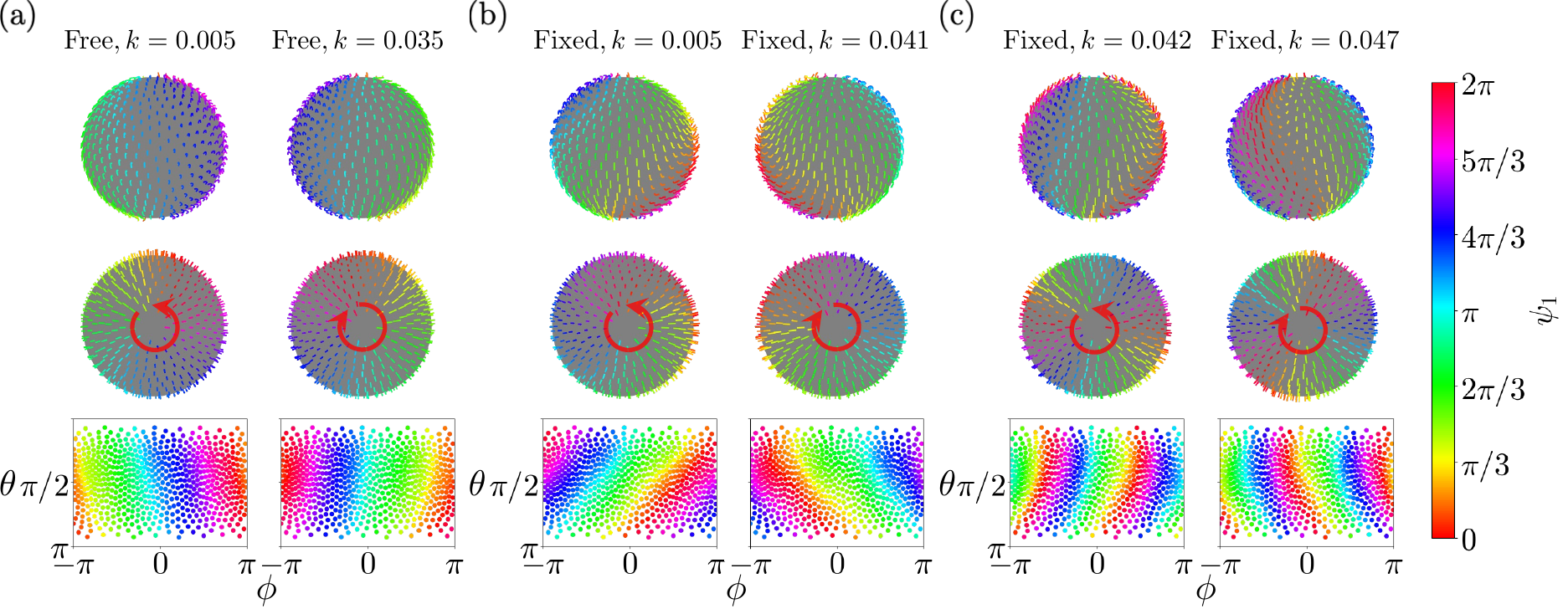}
\end{center}\caption{Diaplectic waves for ciliates with $M=639$ for different values of $k$ and for both held fixed and free-to-swim cases.  For each case, the plots show the phases, $\psi_1$ as a function of cilium surface position in spherical coordinates, where $\theta$ is the polar angle and $\phi$ is the azimuthal angle. Videos showing examples of these waves can be found in the supplementary materials.}
    \label{fig:diaplectic}
\end{figure}

\begin{figure}[h!]
    \centering
    \includegraphics[width=\textwidth]{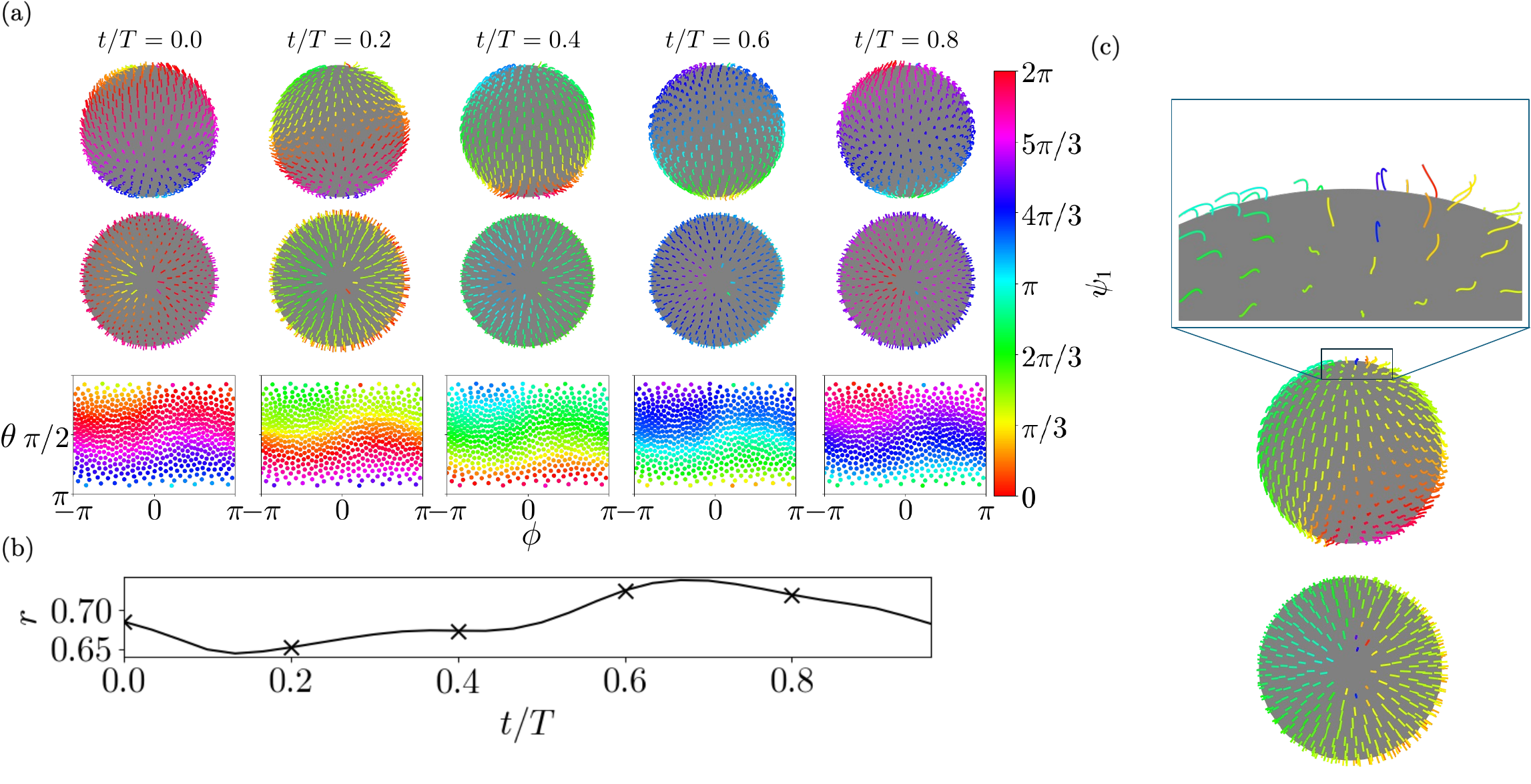}
    \caption{(a)\label{fig:symplectic_ciliate} 
    Time evolution of the cilia phases for a symplectic wave on a free-to-swim ciliate with $M=639$ and $k=0.007$.  In the plots, the phase, $\psi_1$, at each time is shown as a function of cilium surface position in spherical coordinates, where $\theta$ is the polar angle and $\phi$ is the azimuthal angle.  (b)\label{fig:symplectic_period} The Kuramoto order parameter for the symplectic wave over one period.  (c)\label{fig:symplectic_defect} The anterior pole showing the effect of the defect on local coordination. Videos showing the symplectic wave and cilia dynamics near the defect can be found in the supplementary materials.}
    \label{fig:symplectic}
\end{figure}

\begin{figure}[h!]
    \centering
    \includegraphics[width=0.8\textwidth]{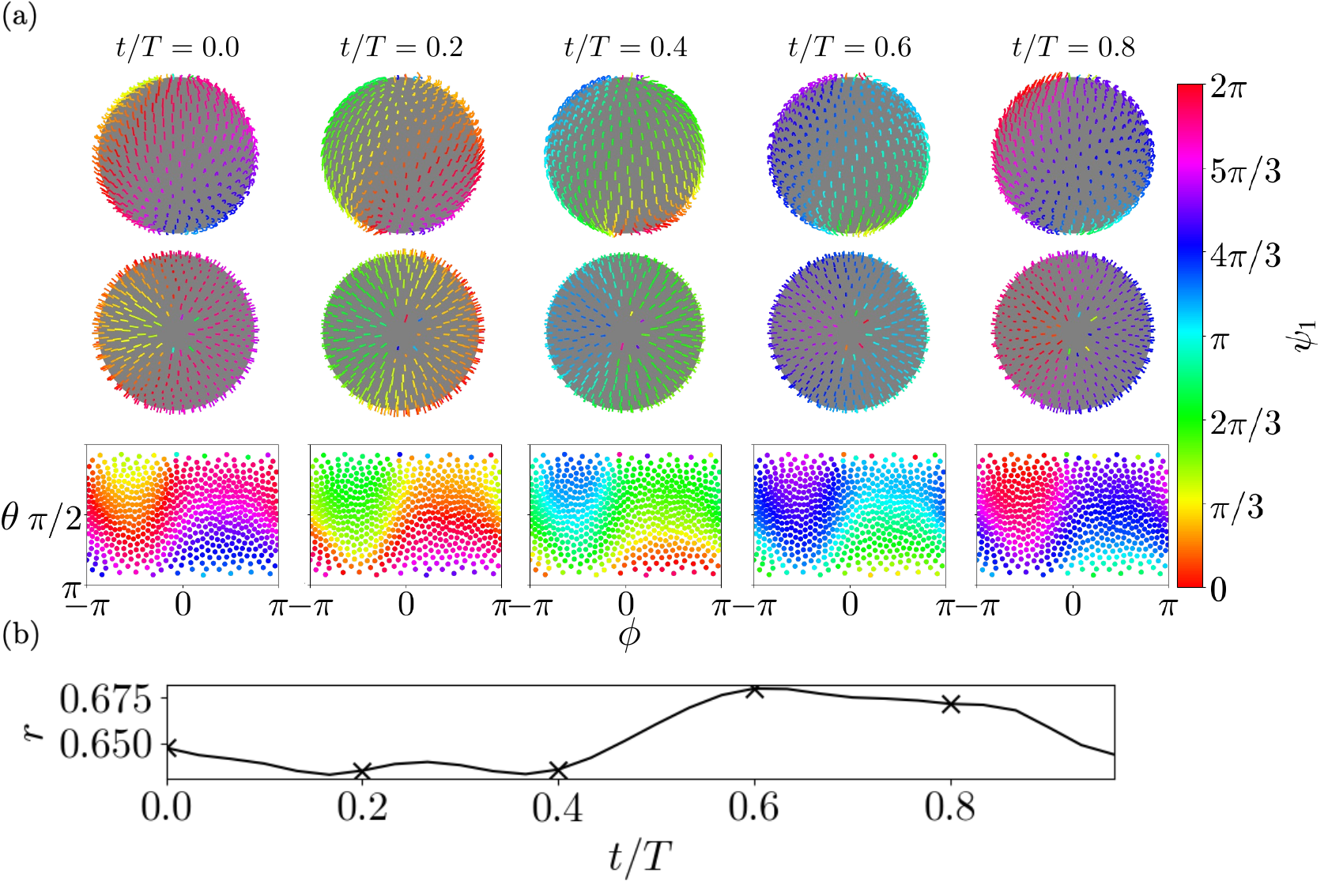}
    \caption{(a)\label{fig:hemisphere_ciliate} Time evolution of a free-to-swim ciliate with $M=639$ and $k=0.017$ after the cilia have reached quasi-symplectic coordination. In the plots, the phase, $\psi_1$, at each time is shown as a function of cilium surface position in spherical coordinates, where $\theta$ is the polar angle and $\phi$ is the azimuthal angle.  (b)\label{fig:hemisphere_period} The Kuramoto order parameter for the symplectic wave over one period.  }
    \label{fig:hemisphere}
\end{figure}

\begin{figure}[h!]
    \centering
    \includegraphics[width=\textwidth]{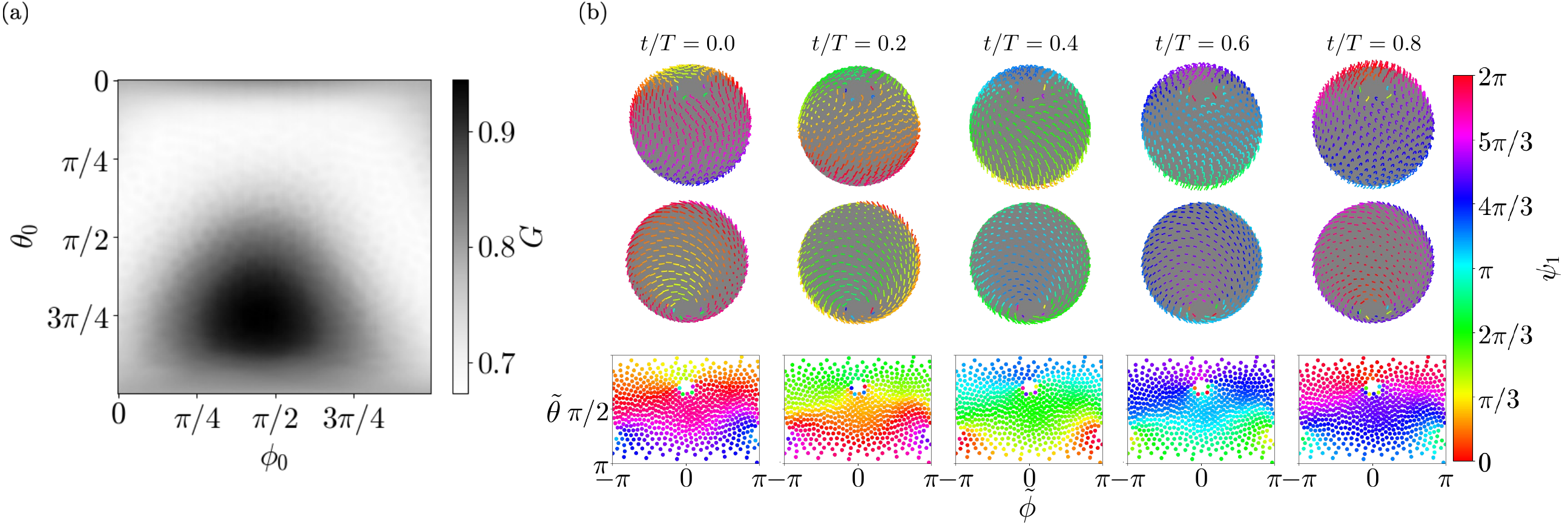}
    \caption{(a)\label{fig:rotated_hemisphere_angle} Heatmap of $G$, indicating the degree of polar propagation of the metachronal wave in the spherical coordinates where $(\theta_0, \phi_0)$ coincides with the anterior pole, $(0,0)$.(b)\label{fig:rotated_hemisphere_ciliate}  Time evolution of the free-to-swim ciliate shown in Fig. \ref{fig:hemisphere} in the rotated coordinate system where $\theta_0 = 1.94$, $\phi_0 = 1.2$, corresponding to the peak value of $G$.   In the plots, the phases, $\psi_1$, at each time are shown as a function of cilium surface position in rotated spherical coordinate system, where $\tilde{\theta}$ is the polar angle and $\tilde{\phi}$ is the azimuthal angle.} 
    \label{fig:rotated_hemisphere}
\end{figure}


We now examine how the long time states depend on the cilium stiffness, $k$, and the condition of the ciliate being free to swim and rotate, or held fixed.  Fig. \ref{fig:order_parameter} shows the values time-averaged order parameter, 
\begin{align}
\langle r \rangle = \frac{1}{10T} \int_{t_f-10T}^{t_f} r(t) dt,
\end{align}
as a function of $k$.  For the free-swimming ciliates, we find that three distinct states emerge at long times, with the bistability appearing below the critical value $k_c < 0.055$.  

First, we find that diaplectic waves arise across the entire range of $k$ that we explored ($0.005 < k < 0.1$).  Fig. \ref{fig:diaplectic}(a) shows two examples of the diaplectic waves for a free-swimming ciliate (see also supplemental material).  We find that the wave shape is largely independent of $k$ and the wavelength at the equator is always $\lambda = 2\pi R$.  Thus, the change in phase as one moves azimuthally around the sphere is $2\pi$.  The wavefront exhibits some variation with the polar angle, but variations in phase are largely in the azimuthal direction, as is the wave propagation.  In addition, due to the symmetry of the spherical surface and the relatively uniform distribution of the cilia, we find that both clockwise and anti-clockwise diaplectic waves can emerge, with the direction of propagation determined by the initial condition.   

Along with the diaplectic waves, at the lowest values of $k$, we also find that symplectic waves emerge for $0.005 \leq k \leq 0.01$, with $\langle r \rangle \approx 0.7$.  The emergence of symplectic waves alongside those that are diaplectic suggests that there is bistability between these two states.  An example of the symplectic state is shown in Fig. \ref{fig:symplectic_ciliate}(a) for $k=0.007$ (see also supplemental material).  The waves initiate very close to the anterior pole and their propagation is almost entirely in the polar direction.  We see, however, that the surface gradient of phase is not completely aligned with the polar direction.  We attribute this to variations in the distribution of the cilia, but perhaps more notably, the presence of the defect at the anterior pole where the cilia beat away from one another.  Examining this area more closely in Fig. \ref{fig:symplectic_ciliate}(c), we see that the cilia in the vicinity of the pole do not appear to coordinate with the rest of the wave, assuming a phase difference relative to that of the nearby wave, as well as each other.  This can also be seen in the corresponding video in the supplemental material.  Thus, the presence of the defect disrupts the otherwise highly coordinated state that emerges.  We also notice that for this state, the value of $\langle r \rangle \approx 0.7$ is rather high.  This is due to the wavelength of the metachronal wave being greater than the size of the sphere, a difference that we explore later in this section.  

Returning to Fig. \ref{fig:order_parameter}, we see that above that for $0.01 < k < 0.055$, we obtain another state whose value of $\langle r \rangle$ increases from $\langle r \rangle \approx 0.6$ at $k = 0.01$ to $\langle r \rangle \approx 0.73$ at $k = 0.055$.  We refer to this state as a quasi-symplectic wave, which, like the symplectic wave described above, appears to be bistable with the diaplectic wave.  An example of this state is shown in Fig. \ref{fig:hemisphere}(a) and supplemental material for $k = 0.017$.  We see that the point of origin of the wave is shifted away from the anterior pole and is accompanied by a change in direction of wave propagation.  Based on this observation, we determine $(\phi^*, \theta^*)$, the point from which the wave originates.  To do this, we first interpolate $\psi_1$ to a grid in new spherical coordinates $(\tilde{\phi}, \tilde{\theta})$ in which the point $\tilde \phi = 0$ and $\tilde \theta = 0$ maps to $ \phi = \phi_0$ and $\theta  =\theta_0$ in the original system.  We then compute numerically the $\tilde \theta$-dependent order parameter,
\begin{align}
    r(\tilde\theta) = \frac{1}{2\pi}\left| \int_0^{2\pi} \exp (i\psi_1(\tilde{\phi}, \tilde{\theta}) ) d\tilde\phi \right|.
\end{align}
Then, to determine $(\phi^*, \theta^*)$, we seek the values of $\phi_0$ and $\theta_0$ that maximise the time-averaged quantity,
\begin{align}
G(\phi_0, \theta_0) = \frac{1}{T} \int^T_0 \int_0^\pi r(\tilde \theta) d\tilde \theta dt. \label{eq:objective_func}
\end{align}
Fig. \ref{fig:rotated_hemisphere}(a) shows $G(\phi_0, \theta_0)$ as a function of $\phi_0$ and $\theta_0$ for the quasi-symplectic state with $k=0.017$ presented in Fig. \ref{fig:hemisphere}.  We can see that $G(\phi_0, \theta_0)$ exhibits a clear maximum at approximately $\theta^* = 1.94$ and $\phi^* = 1.2$.  When we change spherical coordinates such that the pole coincides with this point, we see in Fig. \ref{fig:rotated_hemisphere}(b), that the wave propagation is largely in the polar direction, though there is some disruption when the wave passes over the defect.  From the rotated coordinate system, we also observe the significant increase in coordination after the wave passes the defect, producing the increase in $r$ seen in Fig. \ref{fig:hemisphere}(b) during the second part of the period.  

To better understand the symplectic states, and the limited range of $k$ over which they occur, in Appendix \ref{appendix:planar_k} we consider a single line of cilia on a planar, no-slip surface.  As in the full spherical ciliate simulations, we find that at low values of $k$ symplectic waves emerge.  However, as we increase $k$, the wave transitions from being purely symplectic, to one that is antiplectic.  This transition occurs at values close to $k \approx 0.055$, which was the critical value for symplectic waves in the simulations.  This suggests the possibility that surface curvature restricts the emergence of antiplectic waves.

Broadly speaking, coordination for the ciliate held fixed is similar to that for the freely swimming case, see again Fig. \ref{fig:order_parameter}, though there are some differences to note.  While we find bistability at lower values of $k$, we do not observe the highly symplectic waves with large $\langle r \rangle$, and find only a quasi-symplectic state with values of $\langle r \rangle$ that increase with $k$.  The values of $\langle r \rangle$ for these cases are uniformly lower than those measured for the free-swimming ciliate.  We also see that the range of bistability is reduced for the held fixed case, with the critical value lowered to, $k_c \approx 0.045$.  We again find diaplectic waves with wavelength $\lambda = 2\pi R$ emerge across the entire range of $k$.  Though these waves can also propagate clockwise or anticlockwise, their waveform is different from the one seen for the free-swimming case.  In particular, we see that the wave has a larger gradient in the polar direction, see Fig. \ref{fig:diaplectic}(a).  The main difference with the free-swimming ciliate, however, is that the fixed case exhibits another solution for $k > 0.04$ corresponding to a diaplectic wave with $\lambda = \pi R$ and a lower value of $\langle r \rangle$.  The shape of the wavefront shown in Fig. \ref{fig:diaplectic}(a) and is very similar to the diaplectic wave observed for the free-swimming. 

\subsection{The effect of swimmer size on wavelength} \label{sec:bigspherewave}

\begin{figure}[h!]
    \centering
    \includegraphics[width=\textwidth]{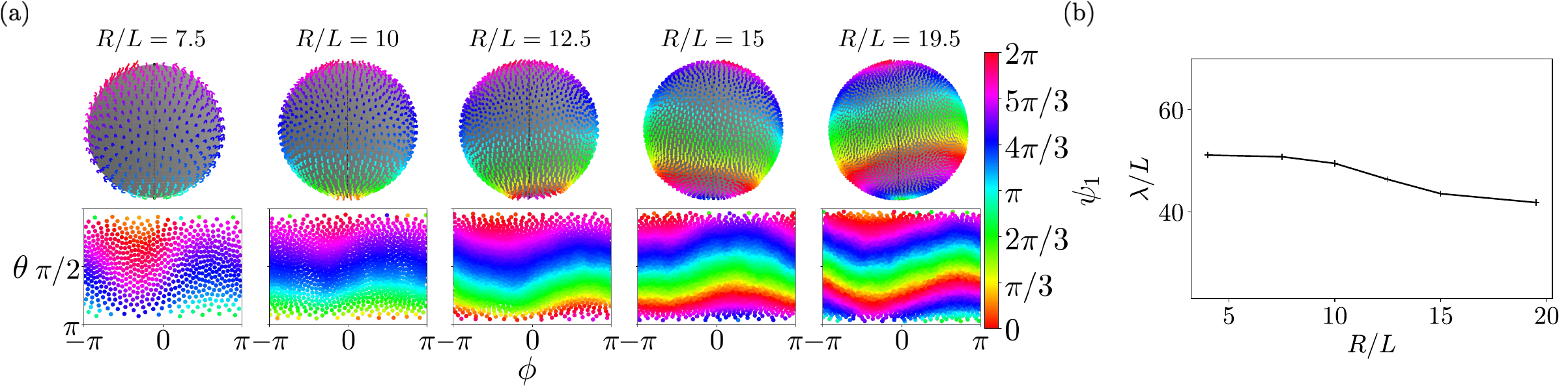}
    \caption{
             (a)\label{fig:bigsphere_ciliate} Symplectic waves on free-to-swim ciliates with increasing $R$ for $k=0.005$ and $ML^2/R^2 = 11.28$.  In the plots, the phase, $\psi_1$, at each time is shown as a function of cilium surface position in spherical coordinates, where $\theta$ is the polar angle and $\phi$ is the azimuthal angle.  (b)\label{fig:bigsphere_wavelength} The wavelength,  $\lambda=2\pi R \Delta \theta/\Delta \psi_1$ as a function of ciliate radius. Videos showing the smallest and largest ciliates can be found in the supplementary materials.}
             
    \label{fig:big_sphere}
\end{figure}

In the previous section, we saw that the symplectic wave has a wavelength greater than the size of the sphere, leading to very high values of $\langle r \rangle$ as the cilia are nearly synchronised.  For diaplectic waves, whether the sphere was held fixed, or allowed to swim freely, the wavelength is set by the size of the sphere, with either one or two complete waves appearing.  For the symplectic wave, however, the mechanisms behind wavelength selection is not as clear.  We extend our investigation of symplectic waves, exploring how they change with the size of the swimmer.  We perform simulations varying swimmer radius $R/L$ from 4 to 19.5, while keeping $k=0.005$ and the cilia number density, $ML^2/R^2 = 11.28$ fixed.  Additionally, we also maintain the same resolution of the spherical surface by keeping $PL^2/R^2 = 728.17$.  Based on these values, $M$ goes from 180 to 4291, and $P$ increases from 11651 to 276888.  For comparison, \textit{Volvox carteri} colonies have , $R/L= 10 - 30$ and the number of somatic cells range from $2000 - 6000$ \cite{pedley2016squirmers}.  
Fig. \ref{fig:big_sphere}(a) shows the emergent symplectic state for the different sized spheres.  
As the sphere increases in size, it is clear that the number of waves also increases.  Plotting the phase as a function of the polar angle $\theta$ allows one to estimate the wavelength for each case, as shown in Fig. \ref{fig:big_sphere}(b).  The resulting estimate as a function of $R/L$ are shown in Fig. \ref{fig:big_sphere}(c).  We see that the wavelength decreases by approximately 20\% from $\lambda \approx 50L$ to $\lambda \approx 40L$ as we increase the swimmer size by nearly a factor of 5.  Thus, in contrast to the diaplectic case, it appears that for the symplectic wave, the wavelength selection mechanism is likely to be set by properties of the interactions between the cilia.  To explore this in more detail, we studied the coordination of cilia pairs and lines on a no-slip planar surface.  This short study is provided in Appendix \ref{appendix:planar_spacing} where we find a natural wavelength to emerge, though shorter than those seen on the spherical surface. Specifically, we obtained  $\lambda/L = 15$ based on the phase shift for a pair of cilia, while we found $\lambda/L = 20$ when we consider a line of cilia.  

\subsection{Beat orientation}


\begin{figure}[h!]
    \centering
    \includegraphics[width=\textwidth]{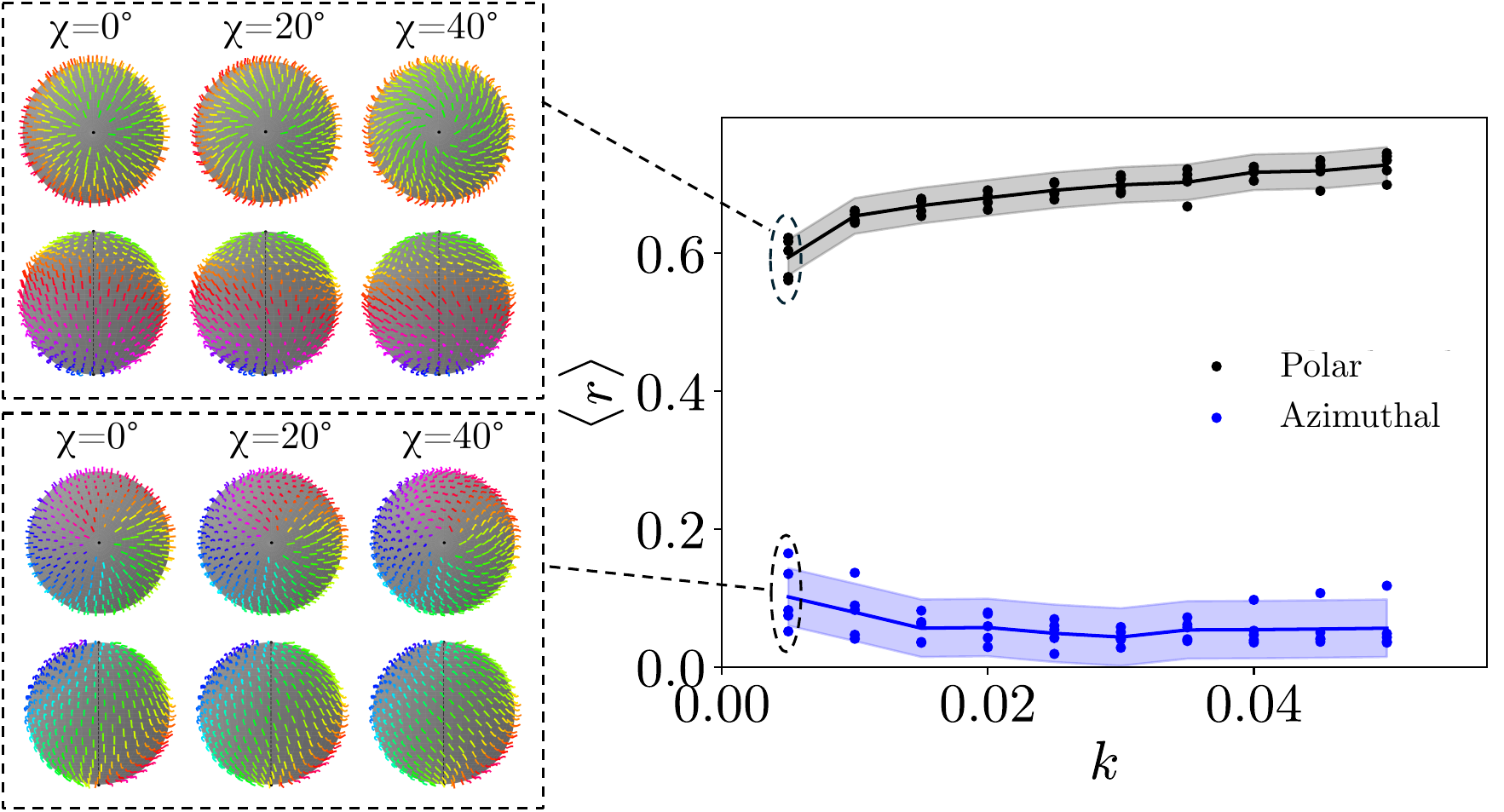}
    \caption{Time-averaged Kuramoto order parameter as a function of $k$ for free-swimming ciliates with different values of beat tilt angle in the range, $0 \leq \chi \leq 40^\circ$.   The emergent waves are largely independent of the beat tilt, as seen from the images of the ciliates in the left panels.}
    \label{fig:tilt}
\end{figure}

A key aspect of ciliate motion, including \textit{Volvox} and \textit{Platynereis} larvae, is that the beat plane of each cilium is tilted relative to the body axis, enabling the ciliate to rotate as it swims.  In our final examination of coordination, we incorporate beat tilt into our simulations by introducing the angle $\chi$, where $\chi = 0 ^\circ$ corresponds to the polar beating, while $\chi = 90^\circ$ corresponds to azimuthal beating.   We again perform simulations run to final times of $1000T$ for swimmers with $M=639$ and $R/L=7.5$, varying $\chi$ from $0^\circ$ to $40^\circ$.  We limit $k$ between $0.005$ and $0.05$ corresponding to the region of bistability for $\chi =0^\circ$ and only consider ciliates that are free to swim.  The resulting values of $\langle r\rangle$ are shown in Fig. \ref{fig:tilt}.  While we observe that there are some small quantitative variations in the exact values of $\langle r \rangle$, comparing the resulting waveforms as shown in Fig. \ref{fig:tilt}, we see that the qualitative features of the coordination are largely unaffected by the tilt, including the fact that both symplectic and diaplectic waves emerge for this range of $k$.  

\section{Propulsion and fluid flows}

Having determined the coordinated states that emerge dynamically, we can assess, compare, and analyse the propulsion and flow fields that these different states generate.  As a result, we can link directly the microscale cilium-level features, such as beat dynamics or stiffness, with the overall functionality of cilia arrays for fluid transport.

    

    

\begin{figure}[h!]
    \centering
    \includegraphics[width=\textwidth]{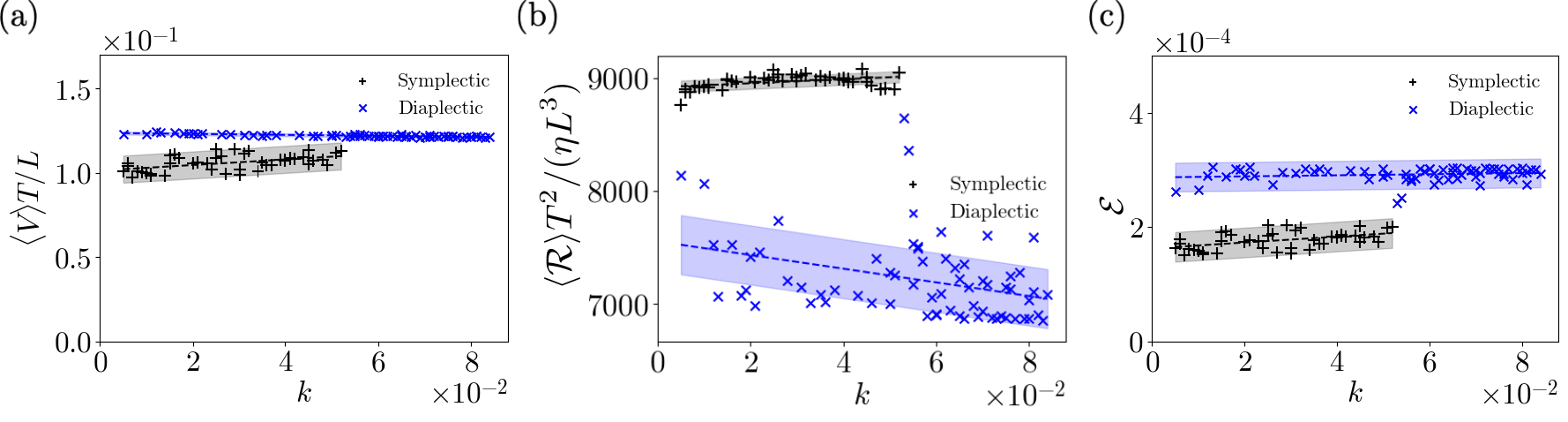}
    \caption{(a) Swimming speed, (b) viscous dissipation and (c) hydrodynamic efficiency for the free-swimming ciliates presented in Fig. \ref{fig:order_parameter}.   Data are grouped according to whether the emergent state is a symplectic-like wave or a diaplectic wave.}
    \label{fig:hydrodynamics}
\end{figure}

\begin{figure}[!h]
    \centering
      \includegraphics[width=1\textwidth]{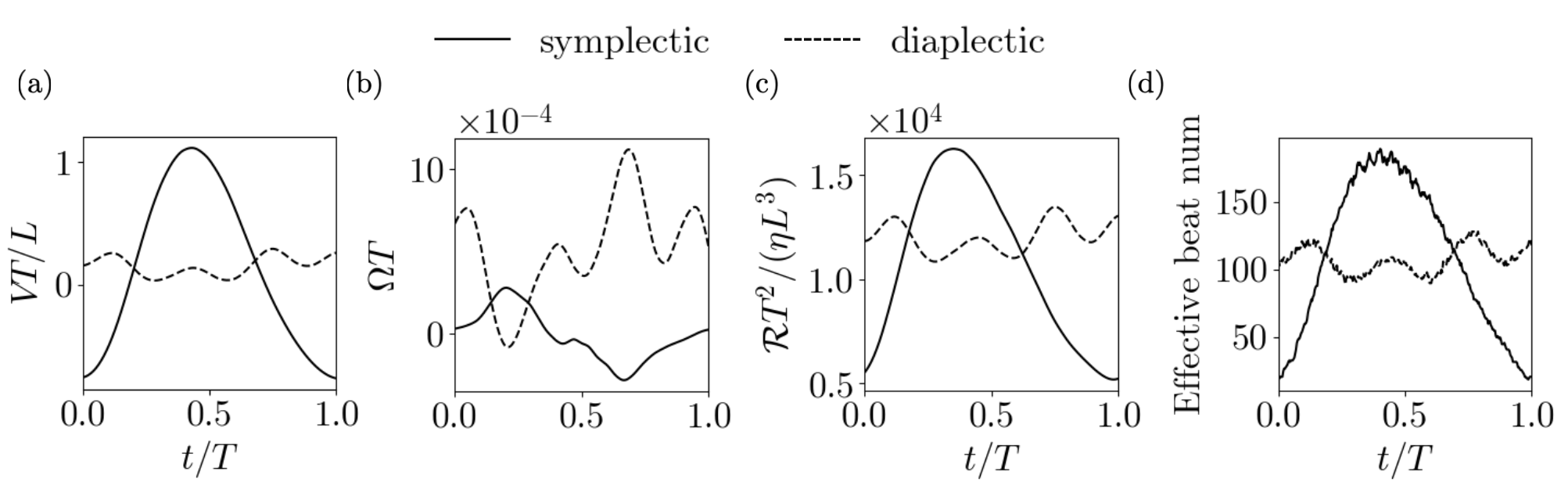}
\caption{(a) Swimming speed, (b) angular speed, (c) viscous dissipation, and (d) number of cilia in the effective stroke, plotted as functions of time over one period, for a ciliate exhibiting symplectic $$(M=639, k=0.005)$$ and diaplectic $$(M=639, k=0.05)$$ metachronal waves.}
\label{fig:hydrodynamics_1T}
\end{figure}

\subsection{Swimming speed, dissipation, and efficiency}

We begin by examining the resulting swimming speed, viscous dissipation, and hydrodynamic efficiency corresponding to the different states achieved as we varied the cilium stiffness.   Fig. \ref{fig:hydrodynamics}(a) shows the period-averaged swimming speed, $\langle V \rangle$ where $V = \bm{V}(t) \cdot \bm{\hat e}(t)$, as a function of $k$ for the emergent states presented in Fig. \ref{fig:order_parameter}(a).  The speeds for the different metachronal waves vary very little with $k$.  The speed for the symplectic state increases by approximately 10\%, while that for diaplectic wave decreases by only 4\%.  What is more pronounced is the difference in speed between the states. We find that the propulsion provided by the diaplectic wave ($\langle V \rangle \approx 0.125 L/T$) is 20\% faster than that of the symplectic wave ($\langle V \rangle \approx 0.105L/T$).   We suspect that the speed difference is not necessarily linked to the metachronal wave pattern, but rather cilia synchrony.  Fig. \ref{fig:hydrodynamics_1T}(a) shows the swimming speed, $V$, as a function of time for symplectic and diaplectic waves and $k = 0.005$.  We see that the symplectic wave produces a velocity that exhibits large temporal variation, and is even negative at one point during the cycle.  The speed for the diaplectic wave case, by comparison, is relatively constant, though smaller oscillations are still present.  The symplectic wave has a higher value of $\langle r \rangle$ meaning that the cilia are nearly synchronised.  Although this allows for high speeds when nearly all cilia are executing their effective stroke, it also produces substantial speed reductions, including reverse motion, when most cilia are in recovery.  Fig. \ref{fig:hydrodynamics_1T}(d) shows the number of cilia in effective stroke as a function of time during the beat, and indeed we see that the peak in the number of cilia in effective stroke coincides with the maximum swimming speed.  For the diaplectic wave, since the wavelength is $\pi R$ or $2\pi R$, at any given moment, the number of cilia in effective and recovery is largely constant, leading to smaller variations and an overall higher swimming speed.

We note that while the speeds are in line with those obtained in similar simulations \cite{Ito2019,Omori2020} with imposed cilia motion, the speeds for both states are significantly lower than measured values of \textit{Volvox}, $\langle V \rangle= 0.20L/T - 0.80L/T$  \cite{pedley2016squirmers}.  We attribute the reduced speed to the Fulford and Blake beat which does not exhibit the very large differences in the effective and recovery strokes that are seen for cilia of swimming microorganisms, including \textit{Volvox} \cite{brumley2014flagellar}.  Indeed, the Fulford and Blake beat was formulated from measurements of respiratory cilium motion, which may instead be tuned for mucus transport rather than aqueous microorganism propulsion.  

Along with the swimming speed, we also examine the hydrodynamic efficiency \cite{lighthill1975,Ito2019},
\begin{align}
\mathcal{E} = \frac{6\pi \eta R \langle V\rangle^2}{\langle \mathcal{R} \rangle},
\end{align}
for the different states as a function of $k$ as shown in  Fig. \ref{fig:hydrodynamics}(c)).  We find that the trends are similar to those observed for the swimming speed, though the difference in efficiency between the symplectic and diaplectic states is more pronounced, with the efficiency being approximately 50\% higher for the diaplectic wave.  While this is, in part, due to the efficiency depending on the square of the swimming speed, we also find that the period-averaged viscous dissipation is also lower for the diaplectic waves (Fig. \ref{fig:hydrodynamics}(b)).  Like the swimming speed, the time-dependence of the viscous dissipation (Fig. \ref{fig:hydrodynamics_1T}(c)) is correlated with the number of cilia in effective stroke.  During the effective stroke, the cilium achieves its maximum velocity during its beat, also exerting maximum stress on the surrounding fluid.  

Finally, Fig. \ref{fig:hydrodynamics_1T}(b) shows the angular speed $\Omega = \bm{\Omega}(t) \cdot \bm{\hat e}(t)$ for these states.  For both symplectic and diaplectic waves there is very limited overall rotation.  This may not be surprising for the symplectic case as the metachronal wave propagates in the polar direction.  For the diaplectic wave, however, the wave propagates azimuthally around the sphere some rotation could be expected.  It is important to recall that although the metachronal wave is in the azimuthal direction, the individual cilia are still beating in the polar direction.  As we show below, the inclusion of beat tilt, which changes the direction of the beat plane, produces the ciliate rotation.

\subsection{Flow fields}

Along with ciliate motion, we also examine the flow fields generated by the symplectic and diaplectic waves.  The flow field over one period for the $k = 0.005$ symplectic case is shown in Fig. \ref{fig:flowfield_symplectic} and in the supplemental material.  The flow remains approximately axisymmetric over the entire period.  For $t/T = 0.2$, when many of the cilia near the midline are in recovery, the streamlines indicate that flow field exhibits a puller-like dipolar structure, where fluid is drawn in toward the ciliate along the $\bm{\hat{e}}$-axis, and ejected laterally.  As time progresses to $t/T = 0.6$, the cilia at the midline are now predominantly in effective stroke, and while we see that the flow remains dipolar, the direction of the flow has reversed and now resembles a pusher-like dipole.  Before returning to the puller-like case, we see that at $t/T = 1.0$, the flow field is closer to that of a neutral swimmer, or force quadrupole (potential dipole), though vestiges of the dipolar field remain.   

\begin{figure}[!h]
    \centering
      \includegraphics[width=1\textwidth]{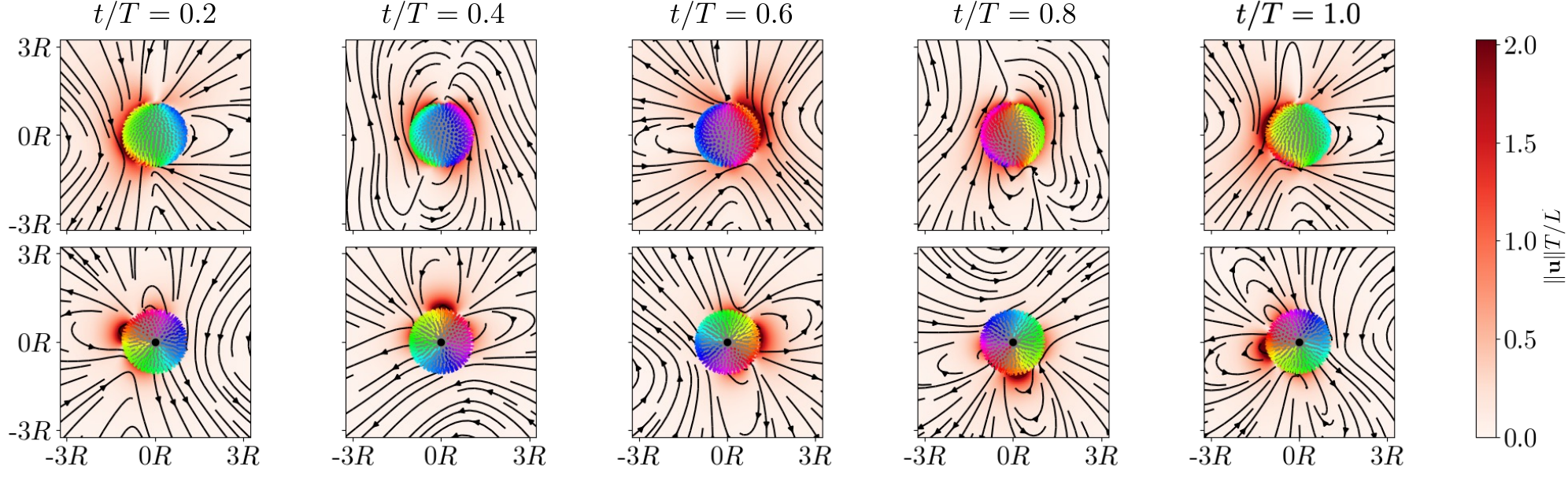}
    \caption{Side (top row) and top (bottom row) views of the flow field at different times for a ciliate exhibiting diaplectic wave coordination corresponding to the case in Fig.\ref{fig:hydrodynamics_1T}.  A video showing the flow field can be found in the supplementary materials.}
\label{fig:flowfield_diaplectic}
\end{figure}

The flow field for the diaplectic case with $k = 0.055$ is shown in Fig. \ref{fig:flowfield_diaplectic} and the supplemental material.  For this case, the flow field is not axisymmetric, with the flow directed towards the posterior in the vicinity of cilia executing effective strokes and toward the anterior on the opposite side of the ciliate where cilia are in recovery.  While the flow is not axisymmetric, we do see that the flow is nearly constant in a frame rotating with the diaplectic wave with some distortions of the streamlines far from the ciliate due to the periodic boundary conditions.

\begin{figure}[!h]
    \centering
      \includegraphics[width=1\textwidth]{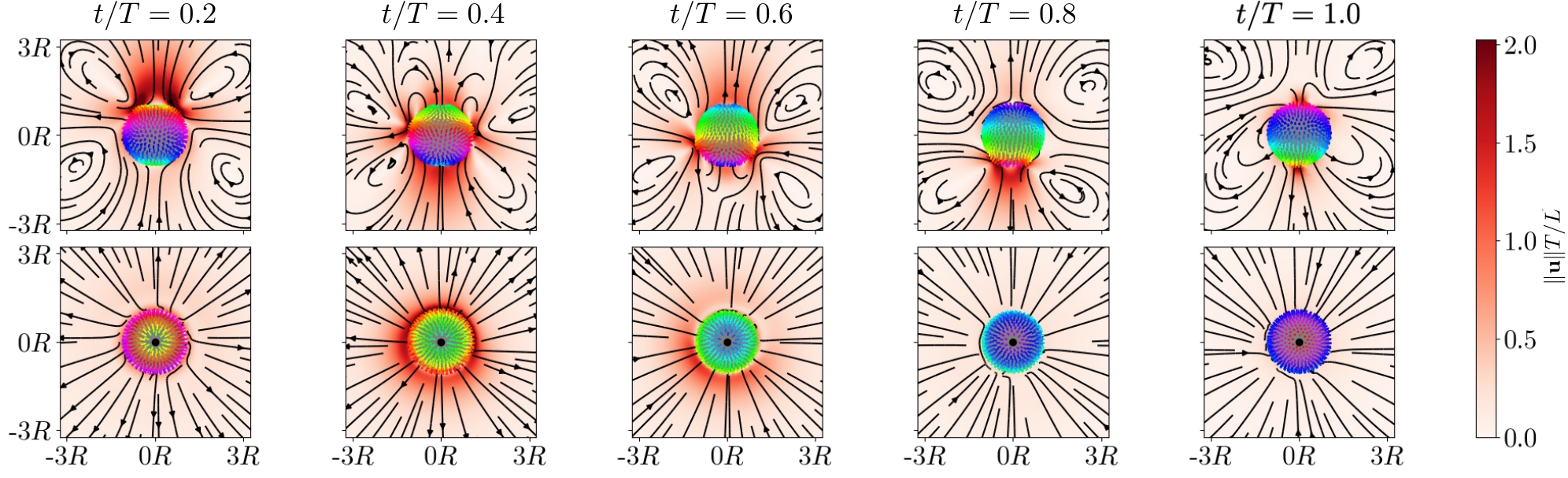}
    \caption{Side (top row) and top (bottom row) views of the flow field at different times for a ciliate with symplectic wave coordination corresponding to the case in Fig.\ref{fig:hydrodynamics_1T}.  A video showing the flow field can be found in the supplementary materials.}
\label{fig:flowfield_symplectic}
\end{figure}

\subsection{Squirming modes}

For the symplectic wave where the flow is nearly axisymmetric, we can analyse the resulting flow by extracting the squirming modes.  Recall that for an axisymmetric squirmer \cite{Blake1971}, the radial and tangential fluid velocities at the squirmer surface are given 
\begin{align}
    \begin{split}
        u_r|_{r=R} &= \sum_n A_n(t) \mathcal{P}_n(\cos\theta) \\
        u_\theta|_{r=R} &=\sum_n B_n(t) \mathcal{V}_n(\cos\theta),\label{eq:squirming}
    \end{split}
\end{align}
where $\mathcal{P}_n$ is the $n$th Legendre polynomial and
\begin{align}
    \mathcal{V}_n(\cos\theta) = \frac{2}{n(n+1)}\sin\theta \mathcal{P}'_n(\cos\theta).
\end{align}
The swimming speed is related to the first two terms through $V = \left(2B_1 - A_1 \right)/3$.

\begin{figure}[!h]
    \centering
      \includegraphics[width=1\textwidth]{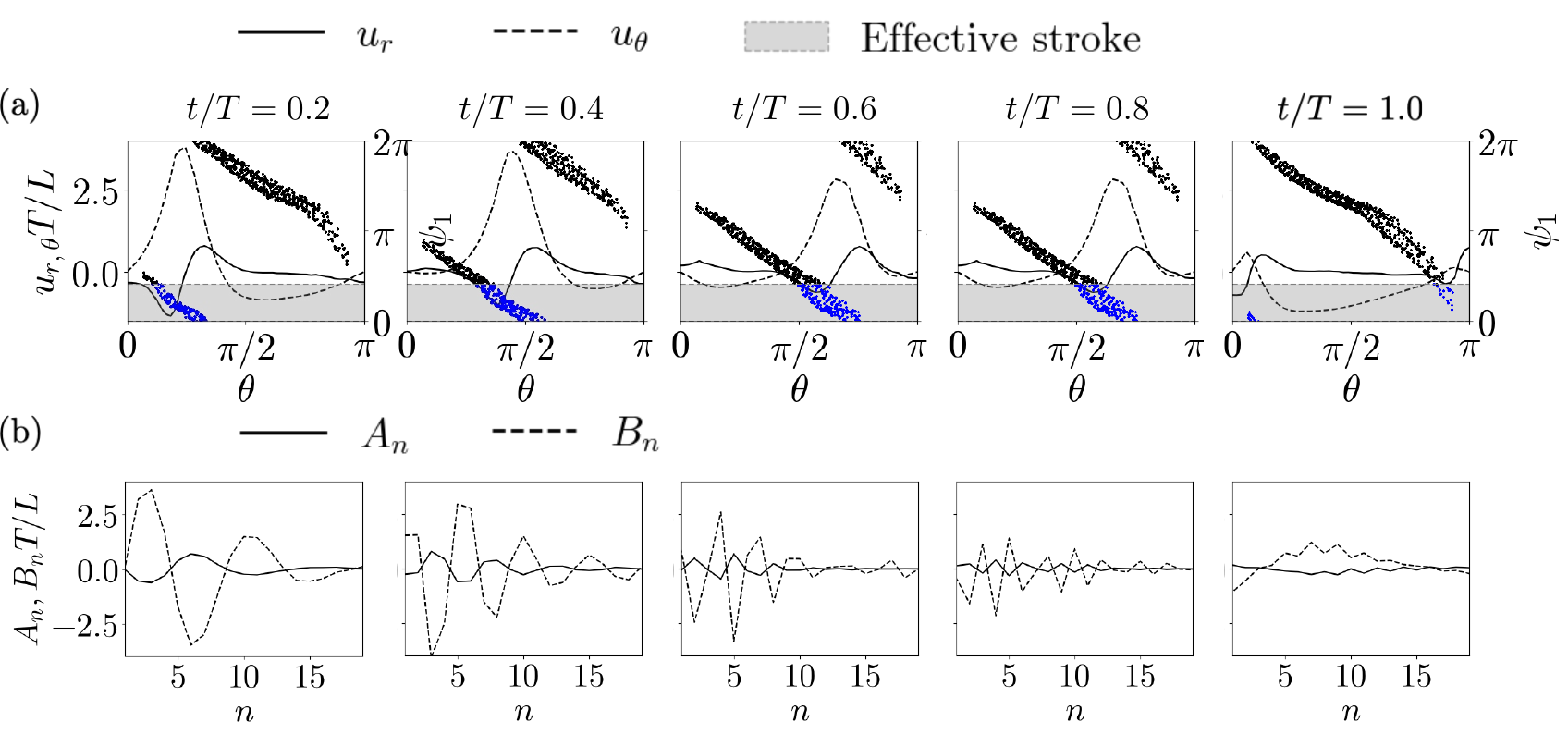}
    \caption{(a) Azimuthally averaged $u_{r,\theta}$ on the envelope surface and $\psi_1$ as a function of polar angle, $\theta$ at different times for a symplectic wave ciliate with $k = 0.005$ corresponding to the case in Fig.\ref{fig:flowfield_symplectic}.   (b) Squirming modes at different times for $u_{r,\theta}$ presented in panel (a). }
\label{fig:u_vs_theta}
\end{figure}

To find the modes, we follow \cite{Brumley2015} and \cite{Ito2019,Omori2020}, and first determine the fluid velocity on a sphere of radius $R_0 = R + 1.1L$ just outside the cilia envelope.  After averaging the fluid velocity in the azimuthal direction, we compute its radial and polar components, $u_{r}(\theta, t)$ and $u_{\theta}(\theta ,t)$, respectively.  Fig. \ref{fig:u_vs_theta} shows $u_{r}(\theta, t)$ and $u_{\theta}(\theta ,t)$ as a function of $\theta$ at different points during the beat period for the $k = 0.005$ symplectic wave.  We see that along with the non-trivial tangential velocity whose peak coincides and propagates with the location of the cilia in effective stroke, we also observe a non-trivial radial flow.  Here, we see that the point where $u_r = 0$ coincides with the location of effective stroke cilia.  This suggests that $u_r$ can be used to quantify metachronal wave propagation, as done in \cite{Brumley2015}.  

\begin{figure}[h!]
    \centering
    \includegraphics[width=0.7\textwidth]{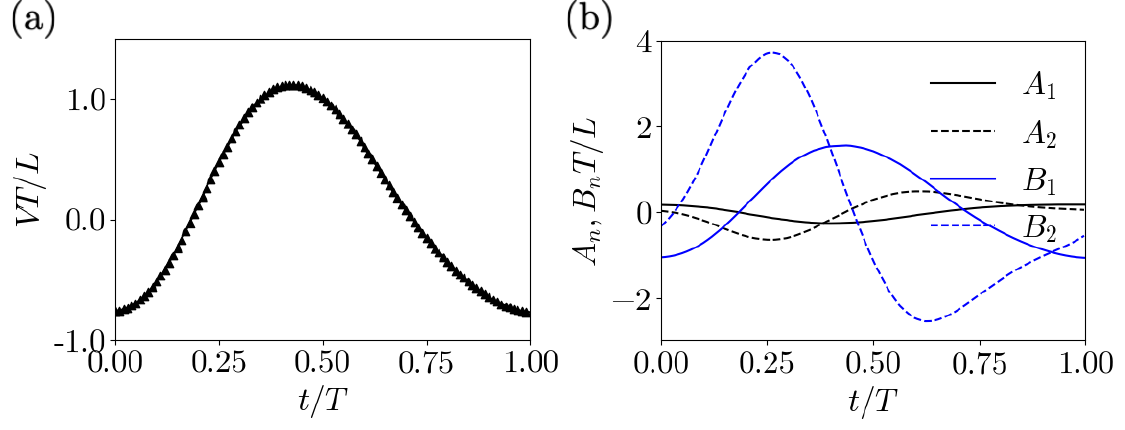}
    \caption{(a) A comparison of the swimming speed given by the simulations (line) and the extracted squirming modes (markers) (b) Time dependence of the lowest squirming modes for the ciliate shown in Fig. \ref{fig:flowfield_symplectic}, with $M=639$, $R/L=7.5$, $k=0.005$.}
    \label{fig:mode_comparison}
\end{figure}

From the $u_{r}(\theta, t)$ and $u_{\theta}(\theta, t)$ on the surface $r = R_0$, we can use the orthogonality of the polynomials $\mathcal{P}_n(\cos\theta)$ and $\mathcal{V}_n(\cos\theta)$ to determine the coefficients $A_n(t)$ and $B_n(t)$ by evaluating numerically the integrals,
\begin{align}
    A_n(t) &= \frac{2n + 1}{2} \int_0^{\pi} u_r(\theta,t) \mathcal{P}_n(\cos\theta) \sin\theta \, d\theta, \\
    B_n(t) &= \frac{1}{8} n(n+1)(2n+1) \int_0^\pi u_\theta(\theta) \mathcal{V}_n(\cos\theta) \sin\theta \, d\theta.
\end{align}
The computed values of $A_n$ and $B_n$ up to $n = 20$ are shown in Fig. \ref{fig:u_vs_theta} at different time during the beat period.  We see that, due to the peaked profile of $u_{r}(\theta, t)$ and $u_{\theta}(\theta, t)$, many modes are necessary to successfully capture all details of the flow field in the vicinity of the ciliate.  As observed in \cite{Ito2019,Omori2020}, using the values of $A_1(t)$ and $B_1(t)$ to determine $V$accurately reproduces the swimming speed from the full simulation (see Fig. \ref{fig:mode_comparison}(a)) with the difference between values being below 1\% for all $t$.  The modes $n=1,2$, as a function of time are shown in Fig. \ref{fig:mode_comparison}(b).  While the $n=1$ modes are linked to the swimming speed, the $n=2$ coefficients instead describe the force-dipole/stresslet\footnote{The dependence of the stresslet on $B_2$ is provided in \cite{ishikawa2006}.  Surprisingly, we could not find the dependence on $A_2$ in the literature.  We determined the dependence on $A_2$ in a straightforward manner, as well as showing no other $A_n$ $n\geq 1$ contribute, following the reciprocal theorem approach described in \cite{lauga2016stresslets}.} associated with the ciliate through 
\begin{align}
\bm{S} = 4\pi\eta a^2 (B_2 - A_2)(\bm{\hat{e}}\bm{\hat{e}}^T - \bm{I}/3).
\end{align} 
We see that $B_2 > 0$ and $A_2 < 0$ for the first half of the cycle yielding a puller-type dipole moment while for the second half of the period, we instead have $B_2 < 0$ and $A_2 > 0$, corresponding to a pusher-type dipole.  These values are consistent with our observations of the flow field in Fig. \ref{fig:flowfield_symplectic}.

\subsection{Ciliate size and beat tilt}

\begin{figure}[h!]
    \centering
    \includegraphics[width=\textwidth]{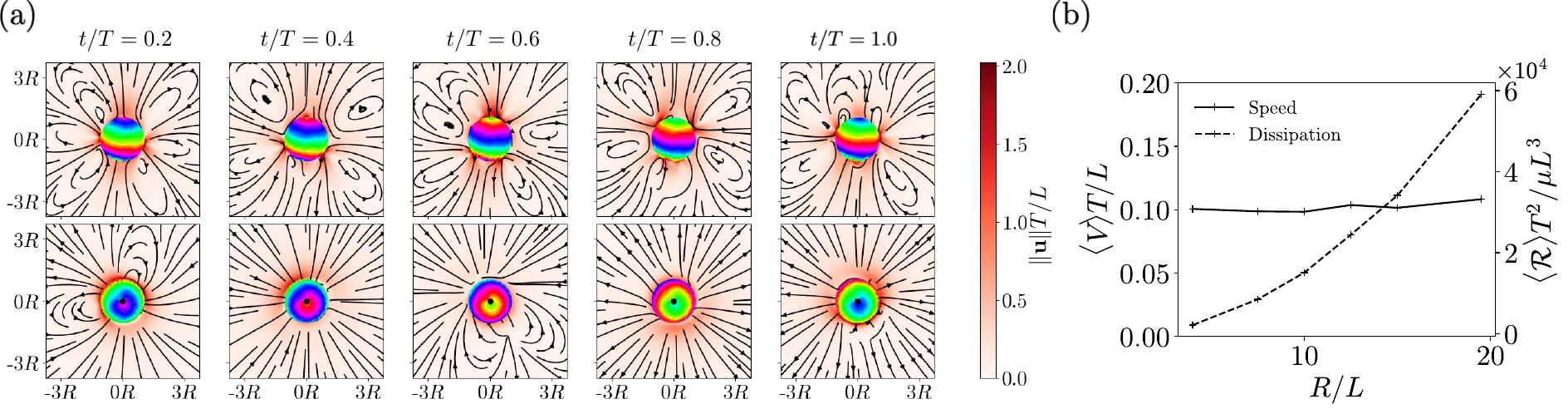}
    \caption{(a) Side (top row) and top (bottom row) views of the flow field at different times for a ciliate with symplectic wave coordination and $R/L=19.5$.  A video showing the flow field can be found in the supplementary materials.  (b)  Speed and viscous dissipation as a function of $R/L$.}
    \label{fig:big_sphere_flowfield}
\end{figure}
In examining symplectic coordination, we saw that as we increased the ciliate size, the number of metachronal waves appearing increases due to its relatively constant wavelength.  This increase in the number of waves leads to changes in the flow field.  As an example, we show in Fig. \ref{fig:big_sphere_flowfield}(a) and the supplemental material the flow field for the ciliate with $R/L = 19.5$ whose symplectic wave has an approximate wavelength, $\lambda = 2\pi R/3$.  At $t/T = 0.2$, we see that as a result of having more waves over the surface, and hence multiple bands of cilia in effective and recovery, the resulting flow resembles a higher-order multipole, with multiple lobes appearing in the streamlines.  At $t/T = 0.4$, the wave has a single effective stroke band just below the equator and the flow field resembles that of a pusher dipole.  At $t/T = 0.8$, the wave has instead a band of cilia in recovery at the equator producing instead a pusher-type dipolar flow field.

Fig. \ref{fig:big_sphere_flowfield}(b) shows the velocity and viscous dissipation for ciliates of increasing size, where we see that the velocity of the swimmer remains constant with the ciliate radius, while the viscous dissipation increases quadratically with $R$.  These dependencies are linked to the fact that the cilia surface density is held constant as we increase $R$, and therefore $M \sim R^2$.   As each cilium is executing the same beat, we expect that the total period-averaged viscous dissipation will scale with the number of cilia $\langle \mathcal{R} \rangle \sim M$, and therefore, $\langle \mathcal{R} \rangle \sim R^2$.  For the swimming speed, keeping the cilia density fixed maintains the same magnitude of the effective surface velocity, thus maintaining the overall speed of the ciliate.  It is interesting to note that the speed does not appear to be affected by the change in the number of metachronal waves on the surface, indicating that the time-average of the surface-averaged velocity remains constant despite the more complicated flows that arise when we increase the swimmer size.  

\begin{figure}[h!]
    \centering
    \includegraphics[width=0.8\textwidth]{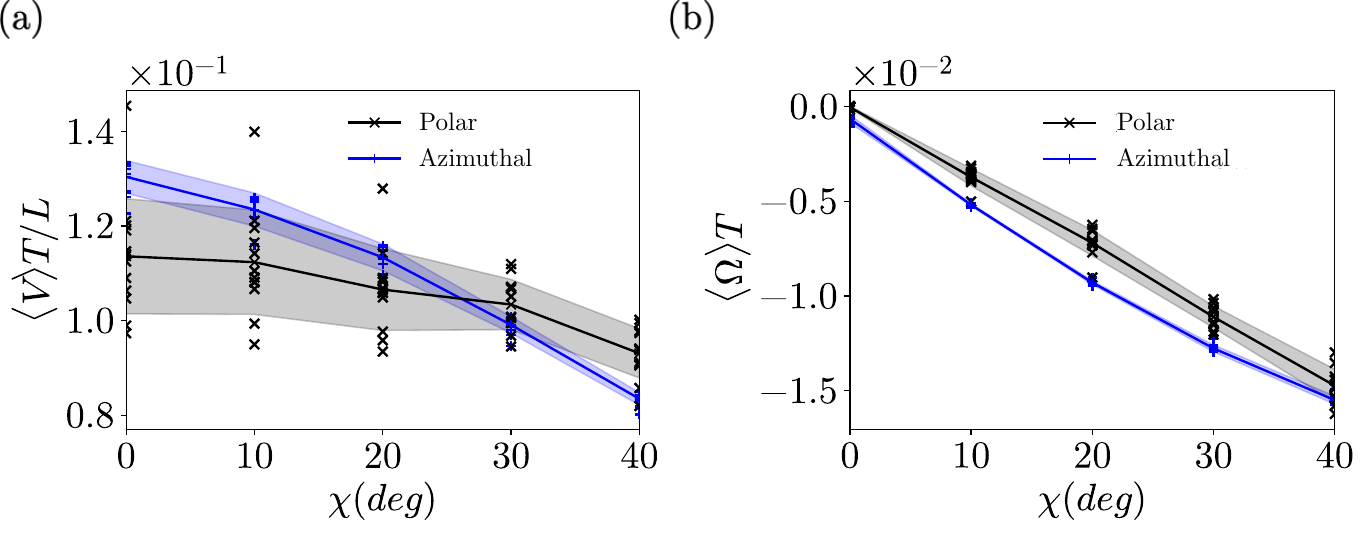}
    \caption{(a) Swimming speed as a function of $\chi$ for polar and azimuthal waves.  (b) Angular speed as a function of $\chi$ for polar and azimuthal waves.  In both panels, the simulations correspond to ciliates with $M=639$, $R/L=7.5$, and $0.005 \leq k \leq 0.05$.}
    \label{fig:tilt_vel}
\end{figure}

Lastly, we explore the effect of cilium beat tilt on ciliate motion and the resulting flow field.  In examining emergent coordination, we saw that tilting the beat plane relative to the body axis did not appreciably affect the polar and azimuthal waves\footnote{When the beat plane is tilted, it is more appropriate to refer to what were the symplectic and diaplectic waves in the zero tilt case as polar and azimuthal waves, respectively.  This is due to the fact that the terms symplectic and diaplectic refer to the direction to propagation relative to the direction of the effective stroke, rather than a direction linked to the underlying surface.} that emerged in the zero tilt case.   Fig. \ref{fig:tilt_vel}(a) shows the time-averaged swimming velocities for the emergent states as a function of the tilt angle, $\chi$.  We see that for both the polar and azimuthal waves the swimming speed decreases with tilt angle.  This is expected as the effective stroke becomes less aligned with the swimming direction as tilt increases.  We do see, however that the speed reduction is more pronounced for the azimuthal wave, and in fact, for the polar wave there is very little change in speed for lower values of $\chi$.  The angular speed for these cases is shown in Fig. \ref{fig:tilt_vel}(b).  Here, we see that the angular speed increases with $\chi$, with a slightly larger increase appearing for the azimuthal wave.  Again, this is expected as the effective stoke has a component in the azimuthal direction, which increases with tilt.  We note that for \textit{Volvox}, the tilt angle is approximately $\chi \approx 15^\circ - 20^\circ$\cite{kirk1998volvox,Brumley2015} and for this tilt, the angular speeds for the model ciliate are similar to the value $\Omega T \approx 0.03 \textrm{rad}$ reported for \textit{Volvox} \cite{Drescher2009}.  

\begin{figure}[!h]
    \centering
      {\includegraphics[width=1\textwidth]{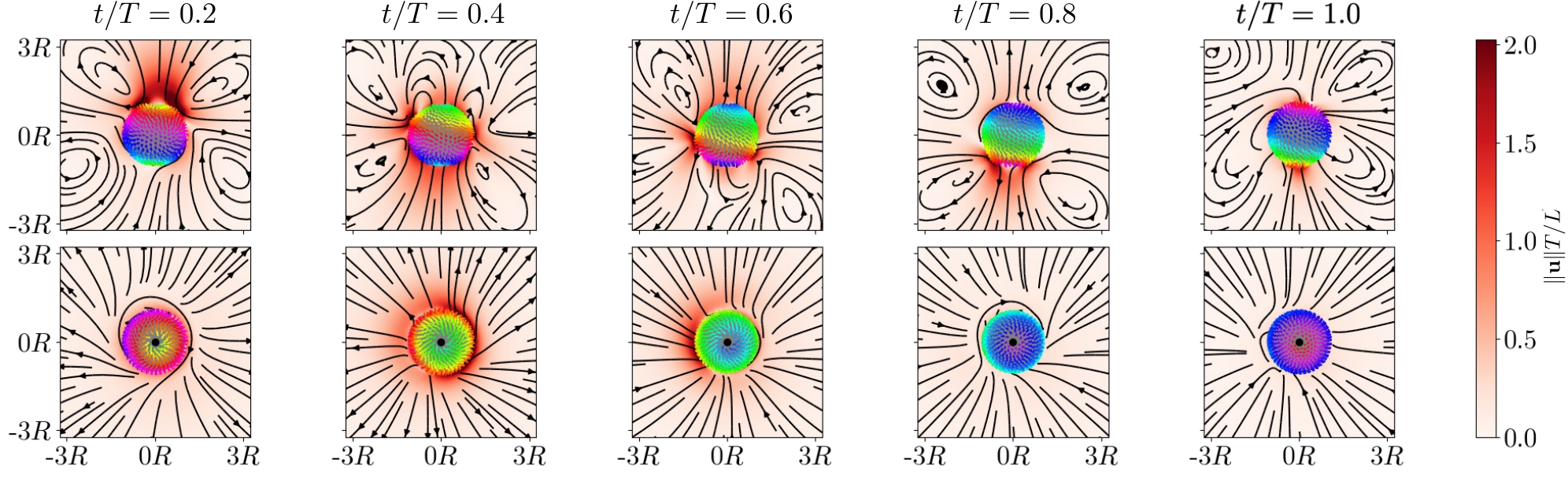}}
    \caption{Side (top row) and top (bottom row) views of the flow field at different times for a ciliate with a polar wave and $M=639$, $R/L=7.5$, $k=0.005$ and $\chi=40^\circ$.}
\label{fig:flowfield_tilt}
\end{figure}

Fig. \ref{fig:flowfield_tilt} and the supplemental material shows the flow field at different times for the polar wave with $\chi=40^\circ$.  From the side view, we see that qualitatively the flow fields are similar to those of the $\chi = 0^\circ$ case shown in Fig. \ref{fig:flowfield_symplectic}, including the switching between puller and pusher states over the course of the period.  Examining the midplane flow from above, however, we see clear differences with the $\chi = 0^\circ$ case.  While far away from the ciliate surface, the flow is radially inward or outward depending whether the ciliate is behaving as a puller or pusher, respectively, closer to the surface, especially when the band of effective strokes passes through the midplane, we see that flow deviates from its radial path and also moves in azimuthal direction.  This flow, which can be attributed to the azimuthal component introduced through beat tilt, is linked to the necessary swirling mode needed to produce torque-free rotation \cite{pedley2016squirmers}.

\section{Discussion}
\label{sec:discuss}
In this study, we performed simulations of a model spherical ciliate by developing the filament oscillator model, a model where each cilium has a filament-like shape, but only two dynamic degrees of freedom.  The model allows for the simultaneous exploration of emergent cilia coordination and the characterisation of the resulting fluid flow.  In performing the simulations, we found an apparent bistability between symplectic and diaplectic waves provided that the cilium stiffness is below the  critical value $k_c \approx 0.055$ for freely swimming ciliates.   This critical value coincided with a transition from sympletic to antiplectic waves in a linear cilia array on a planar, no-slip surface.  Additionally, the wavelength for the symplectic wave appeared to be governed by the cilia interactions, unlike the diaplectic wave, where it was linked with the sphere size.  For free swimming ciliates, the wavelength of the diaplectic wave is $\lambda = 2\pi R$, while for the held fixed case, we can have $\lambda = 2\pi R$ or $\lambda = \pi R$.  In examining the fluid propulsion for these states, we found that the diaplectic waves provided 20\% faster swimming speeds and 50\% greater hydrodynamic efficiencies as compared to the symplectic wave.  We attributed its enhanced propulsion to the shorter wavelength of the diaplectic wave, rather than its direction of propagation.  The wavelength of the symplectic wave is greater than the size of the ciliate leading to the cilia being nearly synchronised.  This results in large temporal oscillations in swimming velocity, including backwards motion.    Interestingly, we did not observe any antiplectic waves on the spherical surface even though such waves have been shown previously \cite{Ito2019} to provide the most efficient swimming for model ciliates with similar geometries.  
Finally, we showed that including beat tilt relative to the body axis did not alter the emergent states, but was critical to introduce ciliate rotation, an important feature that has been linked to microorganism behaviour, such as phototaxis in \textit{Volvox} \cite{drescher2010}.  In addition, for the symplectic waves, beat tilt was able to introduce rotation at a minimal expense to the overall swimming speed. 

Given the spherical shape of the ciliate and density of cilia in our simulations, it is natural to draw comparisons between our results and those reported for $\textit{Volvox}$.  While our model ciliate reproduces symplectic metachronal waves and produces flow patterns similar to those reported for \textit{Volvox} \cite{Brumley2015}, there are also several notable differences.  The first is that the swimming speed for the model ciliate is significantly lower than that reported for \textit{Volvox}.  We believe this difference is in large part due to the differences between the Fulford and Blake beat used in the simulations and the beat of \textit{Volvox} cilia/flagella.  This difference was also noted by \cite{Ito2019,Omori2020}  for their model ciliates using the Fulford and Blake beat with prescribed metachronal waves.   It is interesting to note that the Fulford and Blake beat is derived from measurements of respiratory cilia dynamics \cite{Fulford1986}.  In particular, as compared to \textit{Volvox} flagella dynamics, the Fulford and Blake beat (see Fig. \ref{fig:fil_illustration} above) exhibits less difference between its effective and recovery strokes, a feature essential for effective propulsion at zero Reynolds number.  \textit{Volvox} flagella are relatively straight during their effective stroke and bend quite dramatically during recovery \cite{brumley2014flagellar}.  As a result, as compared to the Fulford and Blakes beat, each individual \textit{Volvox} flagellum is able to provide more propulsion, leading to an overall increase in its swimming speed.  In addition, \textit{Volvox} have pairs of cilia/flagella emanating from their somatic cells, while in our case there is just one.  We suspect that introducing a relative phase between the pair further enhances speeds by reducing the negative impact of the recovery stroke on the swimming speed.

Another notable difference is the presence of the diaplectic wave in the simulations across all parameters in the system, which, to our knowledge, has not been observed for \textit{Volvox}.  Diaplectic waves, however, are indeed common in nature, and have recently been studied in detail for coral larvae \cite{Poon2023}, which are uniformly covered in cilia, and the multicellular larvae of \textit{Platyneries} \cite{poon2025} and the single-celled protist \textit{Didinium} \cite{kourkoulou2025metachronal}, which instead have ciliary bands.  In these examples, however, the wavelength observed is much shorter, $\lambda \sim L \ll R$, than the values of  $\lambda = 2\pi R$ or $\lambda = \pi R$ that we found here for our model ciliate.  The difference in wavelength could again be related to the difference between the cilium beats for these organisms and the Fulford and Blake beat in the simulation, but also the difference in cilium separation.   The cilium separation for other ciliates, such as coral larvae \cite{Poon2023} and \textit{Platynereis} \cite{poon2025}, is $d \ll L$, much smaller than the separation $d \approx L$ used in our simulations.  While hydrodynamic interactions will change with proximity, cilia that are closely separated are also likely to experience steric interactions.  Based on the observations in our study, we presume that the short wavelengths are beneficial to propulsion as at any given moment there will be a significant number of cilia executing their effective strokes.  

While in the particular model ciliate used in this study largely extends from previous, idealised model ciliates, the filament oscillator framework and the larger ciliate construction provides a promising test bed for modelling directly ciliated organisms and querying the role of other physical features in cilia coordination driven and resulting propulsion.  For example, the cilia in our simulations are coupled only through the motion of the surrounding fluid and the motion of the ciliate surface, if it is allowed to move.  For several notable organisms, there are further coupling mechanisms, such as basal coupling, or when cilia density is very high, steric interactions, as discussed in \cite{Wan2024}.   Basal coupling, the direct subsurface coupling of neighbouring cilia through elastic or viscoelastic forces, could come into play, for example, in models of cilia pairs for each somatic cell in \textit{Volvox}, providing a differential, perhaps stronger, coupling between the pairs that works alongside the global coupling provided by hydrodynamics.  The interplay between elastic and hydrodynamic coupling could result in quite different emergent collective dynamics that would be interesting to study.  These additional coupling effects, as well as the propulsion, as well as other arrangement, such as ciliary bands, provide many interesting avenues of research for the filament oscillator model.

\newpage
\appendix

\section{Fulford and Blake Beat Coefficients}\label{app:coeffs}

In our simulations, the cilium beat is provided by the Fulford and Blake \cite{Fulford1986} parameterisation of the cilium movement reported by Sleigh \cite{sleigh1977nature}.  The components of the reference beat \eqref{eq:FBbeat} are of the form \eqref{eq:FBparam}.   For completeness, we reproduce here the values of the coefficients $A_{mn}^{(i)}$  and $B_{mn}^{(i)}$ for $i = 1,2$ in Table \ref{tab:coeffs}.  We note that the $A^{(1)}_{00}$ and $A^{(2)}_{00}$ differ from those in \cite{Fulford1986,Ito2019} by a factor of 2 due to how we have expressed the coefficients in the series.  

\begin{table}[h!]
     \begin{minipage}{.4\linewidth}
       \centering
         \begin{tabular}{ cc|ccccc  }
              \multicolumn{2}{c|}{$A_{mn}^{(1)}$} & \multicolumn{4}{c}{$n$} \\
              & & 0 & 1 & 2 & 3 \\
              \hline
              \multirow{3}{1em}{$m$} & 1  & -0.327 & 0.393 & -0.097 & 0.079 \\
              & 2 & 0.3935 & -1.516 & 0.032 & -0.302 \\
              & 3 & 0.101 & 0.716 & -0.118 & 0.142 \\
         \end{tabular}
         \label{tab:A1_coeffs}
     \end{minipage}%
     \hspace{.05\linewidth}
     \begin{minipage}{.4\linewidth}
         \centering
         \begin{tabular}{ cc|ccccc  }
              \multicolumn{2}{c|}{$A_{mn}^{(2)}$} & \multicolumn{4}{c}{$n$} \\
              & & 0 & 1 & 2 & 3 \\
              \hline
              \multirow{3}{1em}{$m$} & 1  & 0.9475 & -0.018 & 0.158 & 0.01 \\
              & 2 & -0.276 & -0.126 & -0.341 & 0.035 \\
              & 3 & 0.048 & 0.263 & 0.186 & -0.067 \\
         \end{tabular}
     \label{tab:A2_coeffs}
     \end{minipage} \\
     \begin{minipage}{.4\linewidth}
       \centering
         \begin{tabular}{ cc|ccccc  }
              \multicolumn{2}{c|}{$B_{mn}^{(1)}$} & \multicolumn{4}{c}{$n$} \\
              & & 0 & 1 & 2 & 3 \\
              \hline
              \multirow{3}{1em}{$m$} & 1  & 0 & 0.284 & 0.006 & -0.059 \\
              & 2 & 0 & 1.045 & 0.317 & 0.226 \\
              & 3 & 0 & -1.017 & -0.276 & -0.196 \\
         \end{tabular}
         \label{tab:B1_coeffs}
     \end{minipage}%
     \hspace{.05\linewidth}
     \begin{minipage}{.4\linewidth}
         \centering
         \begin{tabular}{ cc|ccccc  }
              \multicolumn{2}{c|}{$B_{mn}^{(2)}$} & \multicolumn{4}{c}{$n$} \\
              & & 0 & 1 & 2 & 3 \\
              \hline
              \multirow{3}{1em}{$m$} & 1  & 0 & 0.192 & -0.05 & 0.012 \\
              & 2 & 0 & -0.499 & 0.423 & 0.138 \\
              & 3 & 0 & 0.339 & -0.327 & -0.114 \\
         \end{tabular}
     \label{tab:B2_coeffs}
     \end{minipage} 
     \caption{Coefficients for the Fulford-Blake beat sequence.}\label{tab:coeffs}
 \end{table}

\section{Saddle point system matrices}\label{appendix:geometric_matrices}

\newcommand*{\horzbar}{\rule[.5ex]{2.5ex}{0.5pt}}

\newcommand{\identity}[0]{\left[ \begin{matrix}
                     1&  &  \\
                     & 1 & \\
                     & & 1 
                 \end{matrix} \right]}

\newcommand{\identityV}[0]{\left[ \begin{matrix}
                     \bm{I}_3 \\
                     \vdots \\
                 \end{matrix} \right]}

\newcommand{\identityH}[0]{\left[ \begin{matrix}
                     \bm{I}_3  &
                     \hdots 
                 \end{matrix} \right]}

\newcommand{\diff}[1]{\left[\begin{matrix}
                        \\
                        \times(\bm{x}_{#1} - \bm{X}) \\
                        \\
                     \end{matrix}\right]
}

\newcommand{\Rmatrix}[1]{\left[\begin{matrix}
                        & \bm{R}_{#1} & \\
                        \\
                     \end{matrix}\right]
}

\newcommand{\RmatrixH}[1]{\left[\begin{matrix}
                        & \bm{R}_{#1} & \hdots \\
                        \\
                     \end{matrix}\right]
}

\newcommand{\RmatrixV}[1]{\left[\begin{matrix}
                        & \bm{R}_{#1} & \\
                        & \vdots \\
                     \end{matrix}\right]
}

\newcommand{\geoK}[2]{\begin{matrix}
                    \vrule \\
                    \bm{K}_{#1_#2} \\
                    \vrule \\
                 \end{matrix}}

\newcommand{\geoKT}[2]{\begin{matrix}
             \horzbar & \bm{K}_{#1_#2}^T & \horzbar
         \end{matrix}}

Appearing in the saddle-point system \eqref{eq:full_saddle_point} are the rectangular matrices $\bm{K}_S$,$\bm{K}_C$, $\bm{\widetilde{K}}_1$, and $\bm{\widetilde{K}}_2$, that relate the rigid body velocity, $\bm{U}$, and cilia phase and orientation velocities, $\bm{\omega}_1$ and $\bm{\omega}_2$, respectively, to the velocities of the surface elements and cilium segments. The matrix $\bm{K}_S \in \mathbb{R}^{3P \times 6}$ is given by
\begin{align}
    \bm{K}_S & = \left[
             \begin{matrix}
                 \bm{I}_3 & 
                  \bm{K}^{1}_S\\
                 \bm{I}_3 & 
                  \bm{K}^{2}_S\\
                  \vdots & \vdots \\
                  \bm{I}_3 & \bm{K}^{P}_S
            \end{matrix}\right],
\end{align}
where $\bm{I}_3$ is the $3\times 3$ identity matrix and 
\begin{align}
    \bm{K}^{p}_S &=\left[\bm{r}_p\right]_\times \in \mathbb{R}^{3\times 3}.
\end{align}
Similarly, the matrix $\bm{K}_C \in \mathbb{R}^{3NM \times 6}$ is given by
\begin{align}
    \bm{K}_C & = \left[
             \begin{matrix}
                 \bm{I}_3 & 
                  \bm{K}^{11}_C\\
                 \bm{I}_3 & 
                  \bm{K}^{12}_C\\
                  \vdots & \vdots \\
                  \bm{I}_3 & \bm{K}^{MN}_C
            \end{matrix}\right],
\end{align}
where $\bm{I}_3$ is the $3\times 3$ identity matrix and 
\begin{align}
    \bm{K}^{nm}_C &=\left[\bm{r}_m +  \bm{R}_{\bm{q}}(\bm{q}_m)\bm{x}_n\left(\psi^{(m)}_1, \psi^{(m)}_2\right)\right]_\times \in \mathbb{R}^{3\times 3},
\end{align}
with $\bm{x}_n = \bm{R}_{\psi}(\psi_2)\bm{\xi}(s_n, \psi_1)$ and the notation, $[\bm{a}]_\times$ to express the $3\times 3$ skew-symmetric matrix such that $[\bm{a}]_\times \bm{b} = \bm{b} \times \bm{a}$, for $\bm{x}, \bm{b} \in \mathbb{R}^3$.
The matrices $\bm{\widetilde{K}}_1 \in \mathbb{R}^{3NM \times M}$ and $\bm{\widetilde{K}}_2 \in \mathbb{R}^{3NM \times M}$ share the structure,
\begin{align}
    \bm{\widetilde{K}}_i = \left[
             \begin{matrix}
                 \bm{\widetilde{K}}^{11}_i \\
                 \vdots \\
                 \bm{\widetilde{K}}^{N1}_i \\ 
                 & \bm{\widetilde{K}}^{12}_i  \\
                 & \vdots \\
                 & \bm{\widetilde{K}}^{N2}_i \\ 
                 \\
                 & & \ddots \\
                 & & & \bm{\widetilde{K}}^{1M}_i  \\
                 & & & \vdots \\
                 & & & \bm{\widetilde{K}}^{NM}_i
            \end{matrix}\right],
\end{align}
for $i= 1,2$ where,
\begin{align}
    \bm{\widetilde{K}}^{nm}_1 &= 
          \bm{R}_{\bm{q}}(\bm{q})\bm{R}_{\bm{q}}(\bm{q}_m)\bm{k}_1(s_n, \psi^{(m)}_1, \psi^{(m)}_2) \in \mathbb{R}^3,\\
      \bm{\widetilde{K}}^{nm}_2 &= \bm{R}_{\bm{q}}(\bm{q})\bm{R}_{\bm{q}}(\bm{q}_m)\bm{k}_2(s_n, \psi^{(m)}_1, \psi^{(m)}_2)\in \mathbb{R}^3,
\end{align}
for $n = 1, \dots N$ and $m = 1, \dots M$.  The expressions for $\bm{k}_1$ and $\bm{k}_2$ are given in \eqref{eq:k1} and \eqref{eq:k2}, respectively.  


\section{Surface element and cilia placement}\label{sec:kmeans}
In constructing the swimmer, we must distribute the surface elements and cilia positions as uniformly as possible over the surface of the sphere. To do this, we utilise spherical $k$-means that we compute using the iterative sequential $k$-means algorithm \cite{MacQueen1967}.  

We first consider the $P$ surface elements.  Applying the sequential $k$-means algorithm involves updating the positions of the points, as well as a set of weights for the points.  Suppose that at iteration $i$, the positions of the elements are $\bm{y}^{i}_1, \bm{y}^{i}_2, \dots, \bm{y}^{i}_P$ and their weights are $w^{i}_1, w^{i}_2, \dots, w^{i}_P$.  The initial positions ($i=0$) are set using the spiral distribution from Saff and Kuijlaars \cite{saff1997distributing} and $w^0_q = 1$ for $q = 1, \dots, P$.   At each iteration, we first generate random vector, $\bm{\xi} = (\xi_1, \xi_2, \xi_3)$, where each component is a Gaussian random variable such that, $\xi_j  \sim \mathcal{N}(0,1)$ for $j = 1,2,3$ and normalise it to obtain $\bm{z} = \bm{\xi}/\lVert \bm{\xi}\rVert$.  After this, the index of nearest surface element is determined via,
\begin{align} 
    p = \operatorname*{argmin}_{q}\lVert\bm{z}- \bm{y}^{i}_q\rVert.
\end{align}
The weights and positions are then updated such that $w^{i+1}_p = w_p^{i}+1$ and $\bm{y}^{i+1}_{p} = \bm{y^*}/\lVert \bm{y^*}\rVert$ where $\bm{y^*} = (w^{i}_p\bm{y}^i_p + \bm{z})/w^{i+1}_p$, and $w^{i+1}_q = w_q^{i}$ and $\bm{y}^{i+1}_{q}=\bm{y}^{i}_{q}$ for $q\neq p$.  We typically repeat the process for $10^7$ iterations, when we find the update procedure no longer produces appreciable changes in the element positions.  

Cilia base positions are determined using the same algorithm, though we also bias placement to avoid regionds near the anterior and posterior poles. We achieve this by requiring that any candidate positions satisfy $\lVert \bm{z} - R \bm{\hat{e}}\rVert \geq 2L$ and $\lVert \bm{z} + R \bm{\hat{e}}\rVert \geq 2L$ for all $k = 1,\dots,K$, where $\hat{\bm{e}}$ is the unit vector along the ciliate body axis, $R$ is the ciliate radius, and $L$ is the cilium length.

\section{Effect of stiffness on the emergent wave}\label{appendix:planar_k}

\begin{figure}[h!]
    \centering
    \includegraphics[width=\textwidth]{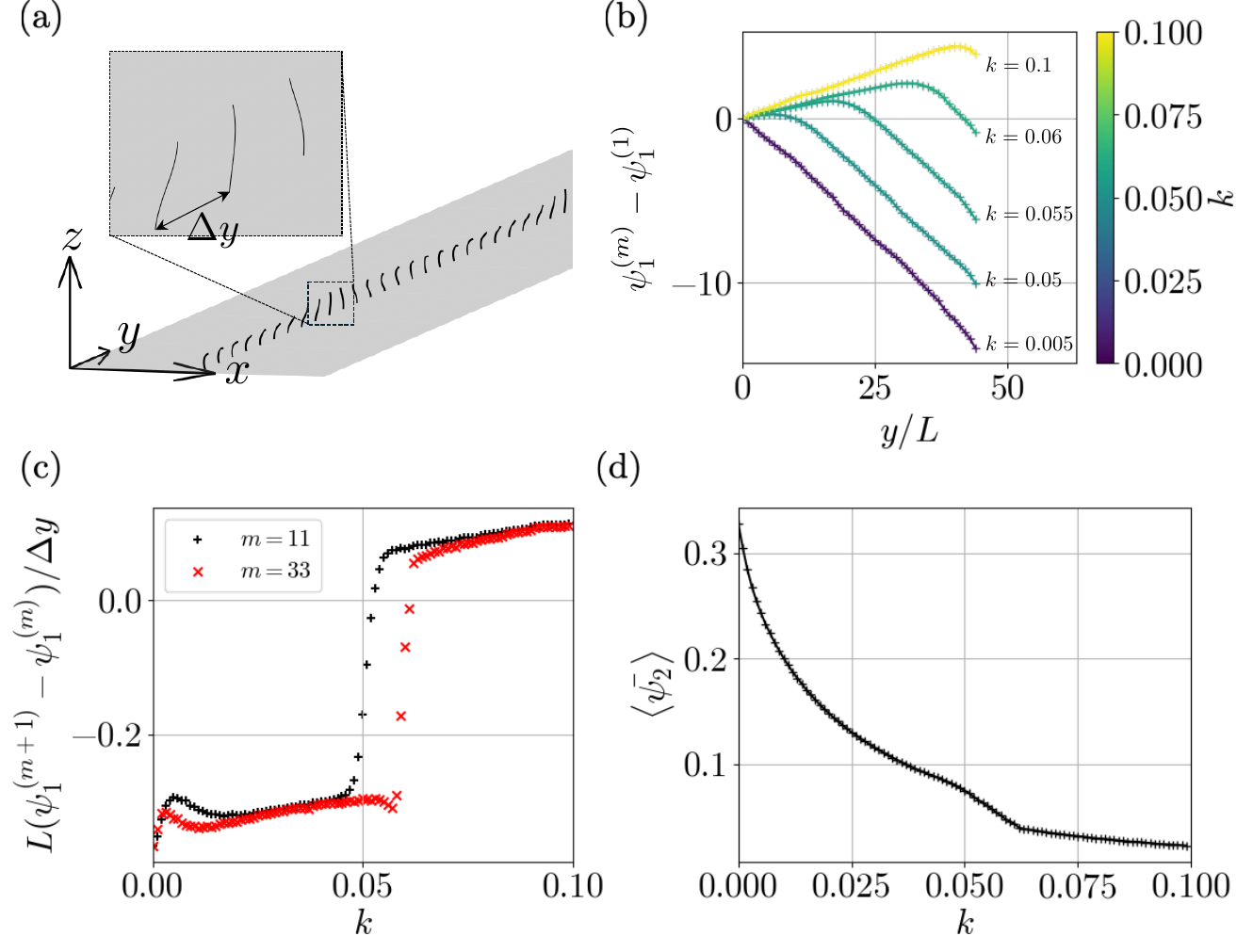}
    \caption{(a) An illustration of the row of cilia from the simulations with  $M=45$ and $\Delta y =L$.  (b) The cilium phases, $\psi^{(m)}_1 -  \psi^{(1)}_1$, for the emergent state as a function of $y$-position for different values of $k$.   (c) Cilium phase gradient, $L(\psi^{(m+1)}_1 -  \psi^{(m)}_1)/\Delta y$, for $m = 11$ and $m = 33$ as a function of spring stiffness $k$.  (d) The spatio-temporal average of $\psi_2$ as a function of $k$.}
    \label{fig:multifil}
\end{figure}

In the ciliate simulations, the symplectic wave emerged only if the stiffness was below a critical value, $k_c \approx 0.055$ for freely swimming ciliates and $k_c \approx 0.045$ for ciliates held fixed.  To better understand this, we consider the simpler arrangement of a single row of $M = 45$ cilia on a planar, no-slip surface, as depicted in Fig. \ref{fig:multifil}(a).  The no-slip condition is achieved by using the RPY-wall mobility matrix \cite{Swan2007} for the hydrodynamic interactions between cilia segments, rather than FCM.  The cilia beat in the $y$-direction and are spaced along the $y$-axis.  To match the average cilia spacing on the spherical surface, the distance between the bases of neighbouring cilia is $\Delta y = L$.  

\begin{figure}[h!]
    \centering
    \includegraphics[width=\textwidth]{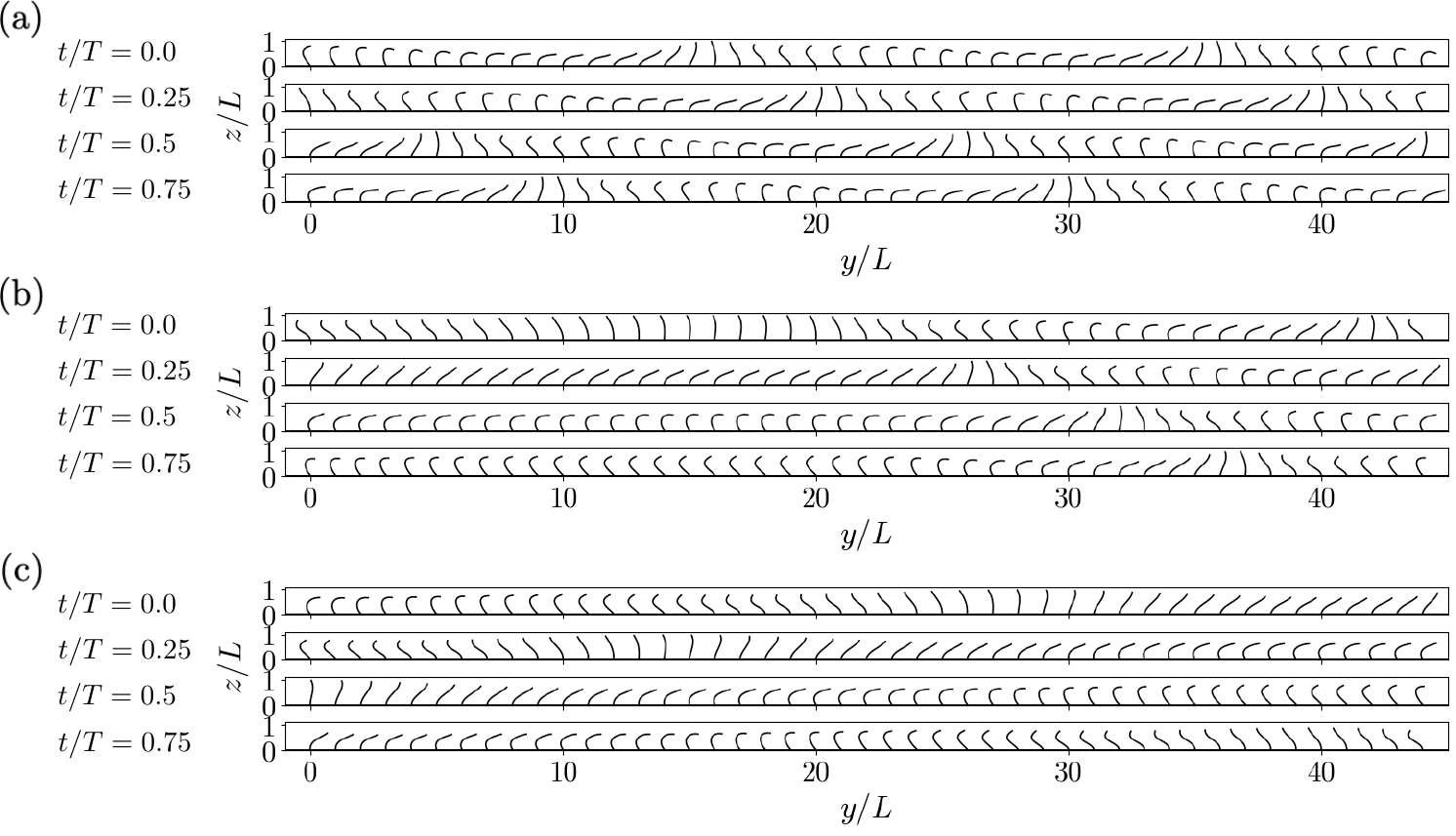}
    \caption{Side view of a row of $M = 45$ cilia with spacing $\Delta y = L$ at different times for (a) $k = 0.005$, (b) $k = 0.055$, and (c) $k = 0.1$. The wave transitions from a short-wavelength symplectic wave ($k = 0.005$) to a long-wavelength antiplectic wave ($k = 0.1$).}
    \label{fig:multifilimage}
\end{figure}

After allowing sufficient time to pass, a coherent state emerges.  The cilium phase, $\psi^{(m)}_1 -  \psi^{(1)}_1$, as a function of $y$-position $y^{(m)} - y^{(1)}$ for this state is shown in Fig. \ref{fig:multifil}(b).  When the stiffness is low, $k = 0.005$, we see that the cilium phase decreases linearly with $y$, corresponding to a symplectic metachronal wave, see Fig. \ref{fig:multifilimage}(a).    As $k$ increases to $k = 0.05$, there is a small region near $y^{(1)}$ where the phase increases with $y$ and by $k = 0.06$, the slope is positive for the majority of the row.  The mixed wave dynamics for the intermediate case $k = 0.055$ are shown in Fig. \ref{fig:multifilimage}(b).  The positive slope corresponds to an antiplectic wave.  By $k=0.1$, the slope for the entire row is positive and the entire wave is antiplectic, see Fig. \ref{fig:multifilimage}(c).   The transition from a symplectic wave to one that is primarily antiplectic occurs in a very narrow range in $k$ centred around $k\approx 0.055$.  Fig. \ref{fig:multifil}(c) shows the slope at $m=11$ and $m=33$, corresponding to positions approximately 1/4 and 3/4 along the row.  The data indicates that at $k\approx 0.05$, 1/4 of the row is propagating an antiplectic wave, while for $k \approx 0.06$, the domain of the antiplectic wave has increased to $3/4$ of the row.  Along with the changes in $\psi_1$, we see that increasing $k$ also results in an overall change in $\psi_2$.  Fig. \ref{fig:multifil}(d) shows the total average (performed over the beat period and all cilia) value of $\psi_2$ as a function of $k$.  As we increase $k$, we see a uniform decrease in $\psi_2$.  The positive values of $\langle \psi_2 \rangle$ observed for lower values of $k$ correspond to an overall rotation of the cilia in the beat direction.  

We see from these simulations that the value of $k$ where the wave transitions from symplectic to antiplectic corresponds to the critical value of $k$ in the full simulations where we no longer see the emergence of symplectic-like waves on the sphere.   We suspect that the curvature of the sphere prohibits the emergence of the antiplectic wave, or instead renders it an unstable state.

\newpage

\section{Coordination of pairs and lines on a planar surface}\label{appendix:planar_spacing}

To better understand the wavelength of the symplectic wave seen in Section \ref{sec:bigspherewave}, we turn again to examine the long time dynamics of cilia on a planar, no-slip surface, considering a pair of cilia, and again a row of $M$ cilia.  

\begin{figure}[h!]
    \centering
    \includegraphics[width=\textwidth]{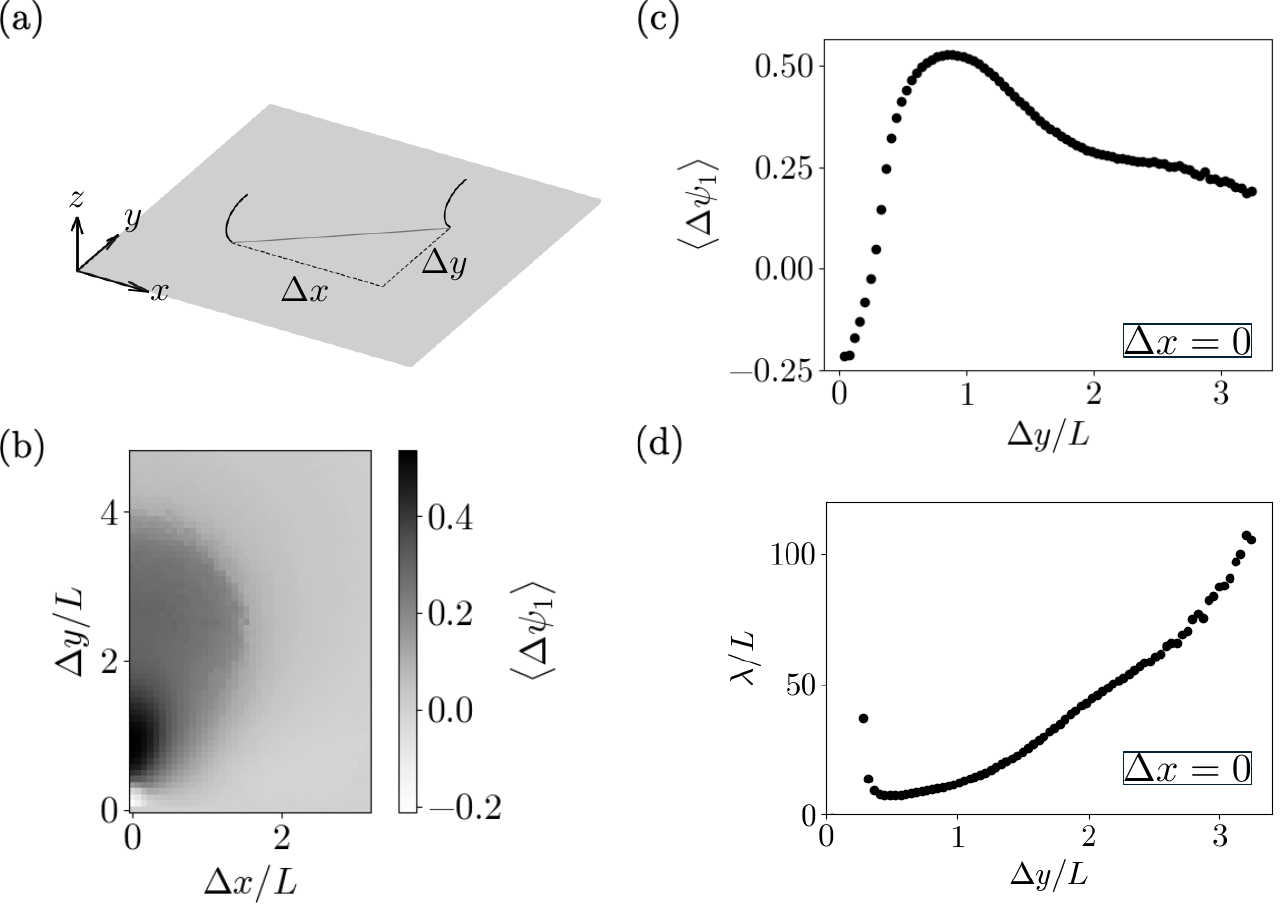}
    \caption{(a) Two filaments placed on a no-slip plane beating in the same direction, with their relative position identified by $\Delta x$ and $\Delta y$. (b) The average phase difference between the two filaments after reaching the equilibrium as a function of relative position, $(\Delta x, \Delta y)$. (c) The average phase difference versus the separation for two filaments that are beating in the same plane. (d) The predicted wavelength of a metachronal wave calculated from the phase difference in (c).}
    \label{fig:twofil}
\end{figure}

The set-up for the pair simulation is shown in Fig. \ref{fig:twofil}(a).  The no-slip surface is at $z = 0 $ and the base of one cilium 1 is located at the origin, while the base of cilium 2 is at $(\Delta x, \Delta y)$.  The cilia are oriented such that their effective strokes are aligned with the $y$-direction and the spring constant is set to $k=0.005$. The cilia are initially in phase and the simulations are run until the phase difference,  $\Delta \psi_1 = \psi^{(2)}_1 - \psi^{(1)}_1$, reaches an asymptotically steady state. The resulting period-averaged phase difference, $\langle \Delta \psi_1 \rangle$, for different values of $\Delta x$ and $\Delta y$ is shown in Fig. \ref{fig:twofil}(b).  We see that when the cilia are separated laterally ($\Delta y =0 $), they remain in phase, while separation in the beat direction results in phase difference.  Fig. \ref{fig:twofil}(c) shows the phase difference for $\Delta x = 0$, where we observe that when the cilia have separation $\Delta y < 0.5L$, the phase shift in negative, while for $\Delta y > 0.5L$ the phase different is positive.  The peak difference of $\langle \Delta \psi_1 \rangle = 0.51$, is found to occur when $\Delta y \approx L$, and decays in magnitude with further separation.  

We can use this phase difference to estimate the wavelength of an emergent wave under the assumption that the pairwise interactions dictate the wavelength such that wavelength of an emergent wave as $\lambda = 2\pi \Delta y/\langle \Delta \psi_1 \rangle$.   The resulting values of $\lambda$ are shown in Fig. \ref{fig:twofil}(d) as a function of $\Delta y$.  Here, we see that the after the phase shift becomes positive, there is an overall increase in the wavelength due to both $\Delta y$ increasing and $\langle \Delta \psi_1 \rangle$ decreasing. In our ciliate simulations, the average cilium separation in the beat direction was $\Delta y \approx L$ and the wavelength decreased from $\lambda = 50L$ to $\lambda = 40L$ as $R$ increased (see Fig. \ref{fig:bigsphere_wavelength}).  We see that the wavelength based on the cilia pair simulations for $\Delta y = L$ is approximately $\lambda \approx 12L$, underestimating the value seen in the full simulation.

\begin{figure}[h!]
    \centering
    \includegraphics[width=\textwidth]{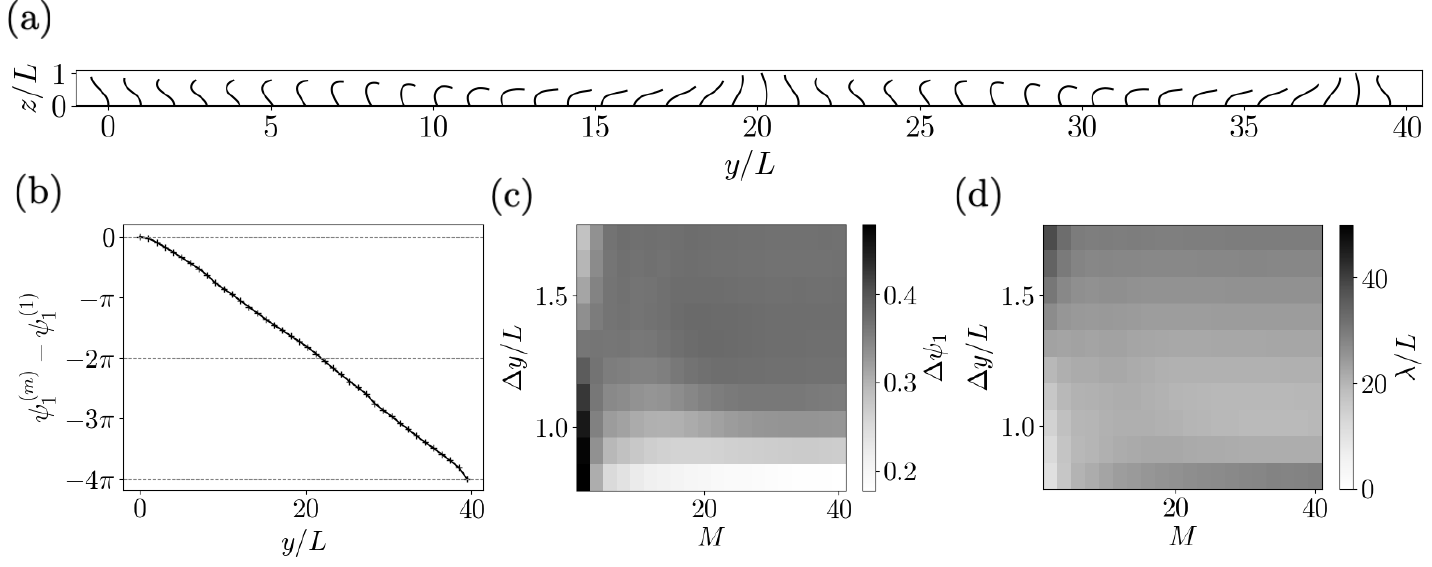}
    \caption{
    (a) An illustration of an array of filaments aligned in a row forming a symplectic metachronal wave, with $M=40$ and $\Delta y =L$.
             (b)\label{fig:multifil_phase_vs_d} Filaments phase as a function of the cilium position, $y$, along the array.
             (c)\label{fig:multifil_phase_heatmap} Phase difference at the centre of the array, $\Delta \psi_1 = \psi^{(M/2+1)}_1 - \psi^{(M/2)}_1$. (d)\label{fig:multifil_phase_wavelength} Wavelength as a function of $\Delta y$ and $M$.}
    \label{fig:multifilwave}
\end{figure}

In addition to pairwise interactions, we also explored the emergent wavelength for a row of cilia on a no-slip, planar surface.  The cilia are uniformly separated by distance $\Delta y$ in the beat direction, as shown in Fig. \ref{fig:multifil}(a) in the previous section.  We examine how the emergent wavelength varies with the number of cilia in the row, as well as $\Delta y$.   One example of such a wave is shown in Fig. \ref{fig:multifilwave}(a) for the case where $M=40$ and $\Delta y = L$.  From these waves, we compute the wavelength, $\lambda$, as well as the average phase difference, $\langle \Delta \psi_1 \rangle$, between neighbouring cilia, which are shown as a function of $\Delta y$ in Fig. \ref{fig:multifilwave}(c) and Fig. \ref{fig:multifilwave}(d), respectively.  We see that the dependence of these quantities on $\Delta y$ for $M \gg 1$ is very different from what we observed for $M = 2$.  Rather than decreasing with $\Delta y$, we instead see that $\langle \Delta \psi_1 \rangle$ increases monotonically with $\Delta y$ and approaches a constant value when $\Delta y > L$.  For the case where $\Delta y =  L$ the wavelength is $\lambda = 20 L$, lower than that seen on the spherical surface, indicating features such as surface curvature, or neighbouring rows are likely to also affect this quantity. 

\newpage

\section{Supplemental Material}
The following movies, generated from the simulations discussed in the
text, are provided as Supplemental Material.
\begin{description}
  \item[Movie S1] Symplectic wave emerging from a random initial
    condition (Fig.~\ref{fig:roadmap}).
  \item[Movie S2] Diaplectic wave emerging from a random initial
    condition (Fig.~\ref{fig:roadmap}).
  \item[Movies S3] Diaplectic wave for a free-to-swim
    ciliate (Fig.~\ref{fig:diaplectic}).
  \item[Movies S4] Diaplectic wave with $\lambda = \pi R$ for a
    held-fixed ciliate (Fig.~\ref{fig:diaplectic}).
  \item[Movie S5] Motion of cilia near the defect near at the anterior pole (Fig.~\ref{fig:symplectic}).
  \item[Movie S6] Symplectic wave for a free-to-swim ciliate
    (Fig.~\ref{fig:symplectic}).
  \item[Movie S7] Symplectic wave for a small ($M=160$)
    free-to-swim ciliate (Fig.~\ref{fig:big_sphere}).
  \item[Movie S8] Symplectic wave for a large ($M=4291$)
    free-to-swim ciliate (Fig.~\ref{fig:big_sphere}).
  \item[Movie S9] Flow field for a symplectic wave
    (Fig.~\ref{fig:flowfield_symplectic}).
  \item[Movie S10] Flow field for a diaplectic wave
    (Fig.~\ref{fig:flowfield_diaplectic}).
  \item[Movie S11] Flow field of the large ciliate with a symplectic
    wave (Fig.~\ref{fig:big_sphere_flowfield}).
\end{description}

\newpage

\nocite{*}

\bibliographystyle{unsrtnat}  
\bibliography{draft}

\end{document}